\documentclass{llncs}

\usepackage[ruled,vlined]{algorithm2e}
\usepackage{stmaryrd}

\newif\ifisdraft
\isdraftfalse

\usepackage{calc}
\usepackage{etex}
\usepackage{times}

\newenvironment{itemize*}%
  {\begin{itemize}%
    \setlength{\itemsep}{3pt}%
    \setlength{\parskip}{0pt}}%
  {\end{itemize}}

\usepackage{amsfonts}
\usepackage{amssymb}
\usepackage{textcomp}
\usepackage{colortbl}
\usepackage{xspace}
\usepackage{alltt}
\usepackage{color}
\usepackage{subfig}
\usepackage{listings}

\usepackage{url}
\usepackage[colorlinks=true,
            urlcolor=blue,
            citecolor=blue,
            linkcolor=blue
            ]{hyperref}

\usepackage{tikz}
\usetikzlibrary{arrows,shapes,shadows}

\usepackage[thmmarks,standard,thref]{ntheorem}
\usepackage{verbatim} 

\usepackage{mathtools}

\usepackage{graphicx}

\usepackage[T1]{fontenc}
\usepackage{enumitem}
\usepackage{fancybox}

\newcommand{\defeq}{\ensuremath{\mathbin{{:}{=}}}}

\ifisdraft
\newcommand{\bl}[1]{\textcolor{red}{\texttt{BL: *** #1 ***}}\marginpar{\textcolor{red}{!!!}}}
\newcommand{\TODO}[1]{\textcolor{blue}{\textbf{TODO:} #1}}
\newcommand{\DZ}[1]{\textcolor{blue}{\textcolor{red}{\textbf DZ}: #1}}
\newcommand{\DZTODO}[1]{\textcolor{blue}{\textcolor{red}{\textbf TODO}: #1}}
\else
\newcommand{\bl}[1]{}
\newcommand{\TODO}[1]{}
\newcommand{\DZ}[1]{}
\newcommand{\DZTODO}[1]{}
\fi

\newcommand{\mypara}[1]{\vspace{5pt}\noindent\textbf{#1.}}
\newcommand{\mysubpara}[1]{\vspace{5pt}\noindent\textit{#1:}}

\newcommand{\Figref}[1]{Figure\,\ref{#1}}
\newcommand{\figref}[1]{Fig.\,\ref{#1}}

\newcommand{\eg}{e.g.,\xspace}
\newcommand{\ie}{i.e.,\xspace}

\newcommand{\rel}[1]{\ensuremath{\mathtt{#1}}}
\newcommand{\mrel}[1]{\mathtt{#1}}

\newcommand{\mskol}[2]{f_{\mathtt{#1}}(#2)}
\newcommand{\pos}[1]{\ensuremath{\mathtt{#1}}}

\newcommand{\glabel}{\ensuremath{\gamma}}

\newcommand{\Gets}{\ensuremath{\mathtt{{:}-}}}
\newcommand{\la}{\ensuremath{\;\Gets\;}}

\newcommand{\idb}[1]{\textit{idb}(#1)}

\newcommand{\KPlus}{\ensuremath{+}\xspace}
\newcommand{\KTimes}{\ensuremath{\times}\xspace}

\newcommand{\mto}{\ensuremath{/}}

\newcommand{\ProvPoly}{\ensuremath{\mathbb{N}[X]}\xspace}
\newcommand{\BoolProvPoly}{\ensuremath{\mathbb{B}[X]}\xspace}

\newcommand{\GLin}{\ensuremath{\mathsf{Lin}}\xspace}

\newcommand{\TrioProv}{\ensuremath{{\mathsf{Trio}(X)}}\xspace}

\newcommand{\RAplus}{\ensuremath{\mathcal{RA}^+}\xspace}

\newcommand{\length}{\ensuremath{\mathrm{len}}\xspace}

\newcommand{\NX}{\ensuremath{{\mathbb{N}[X]}}\xspace}

\newcommand{\gprov}{\ensuremath{\mathrm{\Gamma}}\xspace}
\newcommand{\solved}[1]{\ensuremath{#1^\gamma}}
\newcommand{\gsol}{\ensuremath{\solved{G}}\xspace}
\newcommand{\tj}{\ensuremath{\mathcal{P}}}

\newcommand{\GHow}{\gprov}
\newcommand{\OP}{\ensuremath{\mathrm{\Omega}}}

\newcommand{\datalogneg}{Datalog$^\neg$\xspace}

\newcommand{\WFS}{\ensuremath{\mathcal{W}}\xspace}

\newcommand{\pI}{\ensuremath{\mathrm{I}}\xspace}
\newcommand{\pII}{\ensuremath{\mathrm{II}}\xspace}

\newcommand{\Flr}{\ensuremath{\mathsf{F}}}

\newcommand{\stratI}{\ensuremath{S_{\mathrm{I}}}\xspace}
\newcommand{\stratII}{\ensuremath{S_\mathrm{II}}\xspace}

\newcommand{\Domain}{\ensuremath{\mathtt{d}}\xspace}

\newcommand{\true}{\ensuremath{\mathsf{true}}\xspace}
\newcommand{\false}{\ensuremath{\mathsf{false}}\xspace}
\newcommand{\undef}{\ensuremath{\mathsf{undef}}\xspace}

\newcommand{\green} {{\ensuremath{\textcolor{DarkGreen}{\mathsf{g}}}}}
\newcommand{\red}     {{\ensuremath{\textcolor{DarkRed}{\mathsf{r}}}}}
\newcommand{\yellow}{{\ensuremath{\textcolor{DarkYellow}{\mathsf{y}}}}}

\newcommand{\won} {{\ensuremath{\textcolor{DarkGreen}{\mathsf{W}}}}\xspace}
\newcommand{\lost}     {{\ensuremath{\textcolor{DarkRed}{\mathsf{L}}}}\xspace}
\newcommand{\drawn}{{\ensuremath{\textcolor{DarkYellow}{\mathsf{D}}}}\xspace}

\newcommand{\discup}{\ensuremath{\mathbin{\dot\cup}}}

\definecolor{DarkGreen}{rgb}{0,0.45,0}
\definecolor{DarkRed}{rgb}{0.8,0,0}
\definecolor{DarkYellow}{rgb}{0.6,0.6,0}
\definecolor{DarkGray}{rgb}{0.2,0.2,0.2}

\newcommand{\wonPos}[1]{\textcolor{DarkGreen}{{\emph{\textbf{#1}}}}}
\newcommand{\lostPos}[1]{\textcolor{DarkRed}{{\textbf{\emph{#1}}}}}
\newcommand{\drawnPos}[1]{\textcolor{DarkYellow}{{\emph{\textbf{#1}}}}}

\newcommand{\Q}{\ensuremath{Q}\xspace}  %
\newcommand{\D}{\ensuremath{D}\xspace}  %
\newcommand{\Prog}[1]{\ensuremath{{\Q_{#1}}}\xspace}  %
\newcommand{\ProgD}[2]{\ensuremath{{\Q_{#1},#2}}\xspace}  %

\newcommand{\abc}{\ensuremath{\mathtt{neg}}}
\newcommand{\thop}{\ensuremath{\mathtt{3Hop}}}

\newcommand{\Pabc}{\ensuremath{{\Q_{\abc}}}\xspace}
\newcommand{\PHop}{\ensuremath{{\Q_{\thop}}}\xspace}
\newcommand{\PabcD}{\ensuremath{{\Q_{\abc},\D}}\xspace}
\newcommand{\PHopD}{\ensuremath{{\Q_{\thop},\D}}\xspace}

\newcommand{\etal}{\textit{et al.}\xspace}

\begin{document}

\ifisdraft 
\subsection*{Notation}
\label{sec:notation}

\begin{enumerate}
\item Game $G=(V,M)$ with nodes (aka positions) and edges (aka moves)
\item Query (evaluation) game $G_Q=(V,M)$  (the query evaluation game see above)
\item Play $\pi$: a path, i.e., sequence of edges  in $M$
\item Strategy $S$
  \begin{enumerate}
  \item subset $S\subseteq M$ -- main variant \DZ{Do we want $S$ always be a
  function (eg, deterministic strategy)? I haven't seen where we use the formal
  defition in the paper. As Bertram mentioned in his email, since we use all
  green and all red moves we are the considering mixed strategies instead of
  just the shortest-win / longest-delay.}
  \item mapping $S: V\to V$ -- alternative \DZ{My recommendation: only mention
  strategies on the side and not as integral part of the formal presentation.
  We can later on, mention that the colored edges represent mixed strategies.
  And tidbits like this but do not need the formal treatment here.}
  \end{enumerate}
\item Follower $\Flr(x) = \{ y \mid (x,y) \in M\}$.
  \begin{enumerate}
  \item Reachable $\Flr^+(x)$: nodes reachable from $x$
  \item ... directly or via  the transitive closure $M^+$ of $M$ \DZ{Do you
  intentionally not include reflexive closure here? I assume so.}
  \end{enumerate}
\item Followers reachable via regular expression: $\Flr_R(x)$ \hfill
  e.g. $\Flr_{\green.(\red.\green)^*}(x)$ \\ 
  \DZ{Let's hope we have defined that
  well. Not everybody is as intimately familiar with reg-exp-queries over graphs
  as we are.}
\item Subgraph reachable via regular expression: $G_R(x)$ \hfill  e.g. $G_{\green.(\red.\green)^*}(x)$ 
  \begin{enumerate}
  \item Alternative: $M_R(x)$ \hfill e.g. $M_{\green.(\red.\green)^*}(x)$
  \item NOTE: define $G_R$ via $M_R$; define $\Flr_R$ via $G_R$
  \end{enumerate}
\item Def. \emph{game lineage} $\GLin_G(x) \defeq F_{(\green\mid\red\mid\yellow)^*}(x)$
\item Def. \emph{game provenance} of a game $\gprov \defeq
  G_{(\green\mid\red\mid\yellow)^*}(x) \cup \gamma$ \\
  \DZ{big-cup? Here, looks
  ugly. might be clear later, though}
\item Def. \emph{game provenance} of query evaluation game $\gprov_\ProgD{}{\D}(x) \defeq
  G_{(\green\mid\red\mid\yellow)^*}(x) \cup \gamma$
  \begin{enumerate}
  \item if $\Q$ and $\D$ is clear from context: $\gprov(x)$ 
  \end{enumerate}
\item Solved game $\gsol = (V,M,\gamma)$
\item Solved provenance (query) game $\gsol_\ProgD{}{\D} = (V,M,\gamma)$
  \begin{enumerate}
  \item shorthand: \gsol (if game  $Q$ is clear from the context)
  \item $\lambda$ is nice for labeling L later.%
  \end{enumerate}
\item Game provenance 
\item Query $Q$
\item ABC query $Q_{\rel{FO}}$ %
\item Game query $Q_\rel{G}$ %
\item Query result $Q(D)$ on database $D$
\item Datalog program for $Q$: $P_Q$? needed?
\end{enumerate}

Various forms of query evaluation games have been considered in the
literature \cite{sep-logic-games}, e.g., Hintikka's game-theoretic
semantics GTS \cite{hintikka1996principles}.  The idea of using games
for provenance was inspired more recently, however, by revisiting the
game normal form
\cite{kubierschky1995remisfreie,flum-ICDT-97,flum-TCS-00} for
well-founded \datalogneg~\cite{van1991well}: 

... since every fixpoint query (and thus also every FO query) can be
viewed as a game 
\begin{equation}
\pos{win}(\bar X) \la \pos{move}(\bar X, \bar Y), \neg \pos{win}(\bar Y) \tag{$Q_G$}   
\end{equation}
it is natural to ask: What is the provenance of the game query? 

Can we transfer insights and results from game theory to provenance
of query evaluation by viewing the latter as a game, similar to those
presented in \cite{kubierschky1995remisfreie,flum-ICDT-97}\,?

\TODO{\cite{koehler13:_well_found_proven_games}}

\newpage

\fi

\title{First-Order Provenance Games\thanks{To appear in
    \emph{Peter Buneman Festschrift}, LNCS 8000, 2013.}}

\author{
Sven K\"ohler\inst{1} \and
Bertram Lud{\"a}scher\inst{1} %
\and Daniel Zinn\inst{2} 
}

\institute{
  Dept.\ of Computer Science, 
  University of California, Davis \\
  \email{\{svkoehler,ludaesch\}@ucdavis.edu}
  \and
  LogicBlox, Inc.
  \email{daniel.zinn@logicblox.com}
}

\maketitle

\begin{abstract}
  We propose a new model of provenance, based on a game-theoretic
  approach to query evaluation. First, we study games $G$ in their own
  right, and ask how to explain that a position $x$ in $G$ is won,
  lost, or drawn. The resulting notion of \emph{game provenance} is
  closely related to winning strategies, and excludes from provenance
  all ``bad moves'', i.e., those which unnecessarily allow the
  opponent to improve the outcome of a play.  
In this way, the value of a position is determined by its game provenance.
  We then define \emph{provenance games} by viewing the evaluation of
  a first-order query as a game between two players who argue whether
  a tuple is in the query answer. For \RAplus
  queries, %
  we show that game provenance is equivalent to the most general
  semiring of provenance polynomials \NX.  Variants of our game yield
  other known semirings. %
  However, unlike semiring provenance, game provenance also provides a
  ``built-in'' way to handle negation and thus to answer
  \emph{why-not} questions: In (provenance) games, the reason why
  $x$ is \emph{not} won, is the same as %
  why $x$ is \emph{lost} or \emph{drawn} (the latter is possible for
  games with draws).
  Since first-order provenance games are draw-free, they yield a new
  provenance model that combines \emph{how}- and \emph{why-not}
  provenance.
\end{abstract}

\section{Introduction}

A number of provenance models have been developed in recent years that
aim at explaining {why} and {how} tuples in a query result $\Q(\D)$
are related to tuples in the input database \D (see
\cite{cheney2009provenance,grigoris-tj-simgodrec-2012} for recent
surveys).  Motivated by applications in data warehousing, Cui
\etal~\cite{cui2000tracing} defined a notion of data \emph{lineage} to
trace backward which tuples in $D$ contributed to the result.
Buneman \etal~\cite{buneman01:_why_where} refined and formalized new
forms of \emph{why}- and \emph{where}-provenance, and introduced a
notion of (minimal) witness basis to do so.  Later, Green \etal
\cite{green2007provenance} proposed a form of \emph{how}-provenance
through \emph{provenance semirings} that emerged as an elegant,
unifying framework for provenance.  For \RAplus (positive relational
algebra) queries, provenance semirings form a hierarchy
\cite{green2011containment}, with \emph{provenance polynomials}
\ProvPoly as the most informative semiring at the top (i.e., providing
the most detailed account \emph{how} a result was derived), and other
semirings with ``coarser'' provenance information below, e.g.,
\emph{Boolean provenance polynomials} \BoolProvPoly
\cite{green2011containment}, \emph{Trio} provenance
\cite{benjelloun2006uldbs},
\emph{why}-provenance~\cite{buneman01:_why_where}, and \emph{lineage}
\cite{cui2000tracing}.
The key idea of the unifying framework is to {annotate} each tuple in
the input database \D with an element from a semiring $K$ and then
propagate $K$-annotations through query evaluation. Semiring-style
provenance support has been added to practical systems, e.g.,
\textsc{Orchestra}~\cite{green2007update} and
\textsc{Logic\-Blox}~\cite{huang2011datalog}.
However, the semiring approach does not extend easily to negation and
other non-monotonic constructs, thus spawning further research
\cite{geerts2010database,green2011reconcilable,amsterdamer2011provenance,amsterdamer11:_limit_of_proven_for_queries_with_differ}.

In this paper, we take a fresh look at provenance by employing
\emph{games}.  Game theory has a long history and many applications,
e.g., in logic, computer science, biology, and economics.  The first
formal theorem in the theory of games was published by Ernst Zermelo
exactly 100 years ago
\cite{zermelo13:_ueber_anwen_mengen_theor_schac}.\footnote{Some
  confusion prevails about Zermelo's theorem, but it is all sorted out
  in \cite{schwalbe2001zermelo}.}  In 1928, von Neumann's paper
``\emph{Zur Theorie der Gesellschaftsspiele}''
\cite{neumann28:_zur_theor_gesel} marked the beginning of game theory
as a field. In it he asks (and answers) the question of how a player
should move to achieve a good outcome.  We employ such ``good'' moves
to define a natural notion of provenance for games $G$, which we call
\emph{game provenance} $\gprov\,({=}\,\gprov_G)$, and which is thus
closely related to \emph{winning strategies}.
The crux is that by considering only ``good'' moves while ignoring
``bad'' ones, one can get a game-theoretic {explanation} for why a
position is {won}, {lost}, or {drawn}. By viewing query evaluation as
a game, we can then apply game provenance to obtain an elegant new
provenance approach, which we call \emph{provenance games}.

\mypara{Game Plan} In Section~\ref{sec-general-games} we introduce
basic concepts and terminology for games $G$ and show how to solve
them using a form of backward induction.
We then discuss the regular structure inherent in solved games \gsol
and use it to define our notion of game provenance \gprov. The solved
positions imply a labeling of moves as ``good'' or ``bad'', which we
then use to define the game provenance $\gprov(x)$ of position $x$ as
the subgraph of $G$, reachable from $x$ without ``bad'' moves.
The value of a position is determined by its game provenance, and it
captures why and how a position is won, lost, or
drawn. %

In Section~\ref{sec-provenance-games} we propose to apply game
provenance to first-order (FO) queries in \datalogneg form, by viewing
the evaluation of query $Q$ on database $D$ as a game $G_{Q,D}$.  By
construction, our \emph{provenance games} yield the standard semantics
for FO queries.  For positive relational queries \RAplus, game
provenance $\gprov_{Q,D}$ is equivalent to the most general semiring
of provenance polynomials \NX.  Variations of the provenance game
yield other semirings, e.g., %
\TrioProv.  While our provenance games are equivalent to provenance
semirings for positive queries, the former also handle negation
seamlessly, as complementary claims and negation are inherent in
games.  Provenance games can thus also answer \emph{why-not} questions
easily: The explanation for why $x$ is \emph{not} won is the same as
why $x$ is lost (or drawn, for games that are not draw-free). Since
provenance games are always draw-free for first-order queries, we
obtain a simple and elegant provenance model for FO that combines
how-provenance and why-not provenance.  In
Section~\ref{sec-conclusions} we conclude and suggest some future
work.

\section{Games}\label{sec-general-games}

We consider games as graphs $G=(V,M)$, where two players move
alternately between \emph{positions} $V$ along the edges
(\emph{moves}) $M\subseteq
V\times V$.  We assume that $G$ is finite, i.e., $|V| <
\infty$,\footnote{Many game-theoretic notions and results carry over
  to the transfinite case; cf.\ \cite{flum2000games}.}
but game graphs can have cycles and thus may result in infinite
plays.
Each $v_0\in V$ defines a {game} $G^{v_0} = (V, M, v_0)$
starting at position $v_0$. 

A \emph{play} $\pi$ (= $\pi_{v_0}$) of $G^{v_0}$ is a (finite or infinite)
sequence of edges from $M$:
\begin{equation}
  v_0\stackrel{M}{\to}  v_1
  \stackrel{M}{\to}  v_2  
\stackrel{M}{\to} \cdots 
 \tag{$\pi$}
\end{equation}
i.e.,  where for all $i=0, 1,2, \dots$ the edge $v_{i} \stackrel{M}{\to} v_{i+1}$ is a
 move $(v_{i},v_{i+1})\in M$.  A play $\pi$ is \emph{complete}, either
 if it is infinite, or if it ends after $n=|\pi|$ moves in a sink of
 the game graph.
The player who cannot move {loses} the play $\pi$,
while the previous player (who made the last possible move) {wins} $\pi$. 
Thus, if $|\pi|=2k+1$, we have  $\pi = $
\begin{equation}
  v_0\stackrel{\pI}{\to} v_1 \stackrel{\pII}{\to} v_2 \stackrel{\pI}{\to} \cdots \stackrel{\pII}{\to} v_{2k} \stackrel{\pI}{\to} v_{2k+1}    
\tag{\pI moves last}
\end{equation}
and  $\pi$ is \emph{won} for \pI. Conversely, if \pII
moves last, then  $|\pi|=2k$ for some  $\pi=$
\begin{equation}\label{play-ii}  
  v_0\stackrel{\pI}{\to} v_1 \stackrel{\pII}{\to} v_2
  \stackrel{\pI}{\to} \cdots \stackrel{\pII}{\to} v_{2k}
 \tag{\pII moves last}
\end{equation}
so $\pi$ is  \emph{lost} for \pI, and \pII wins the
play. A play $\pi$ of infinite length is a \emph{draw} (in
\emph{finite} games $G$, this means that $M$ must have a cycle).

\begin{figure}[t]
  \centering
~\hfill
\subfloat[What are the ``good moves'', e.g., in position \texttt{e}?
Is \texttt{e} won (or lost, or drawn), and if so how?]{
  \includegraphics[width=.36\columnwidth]{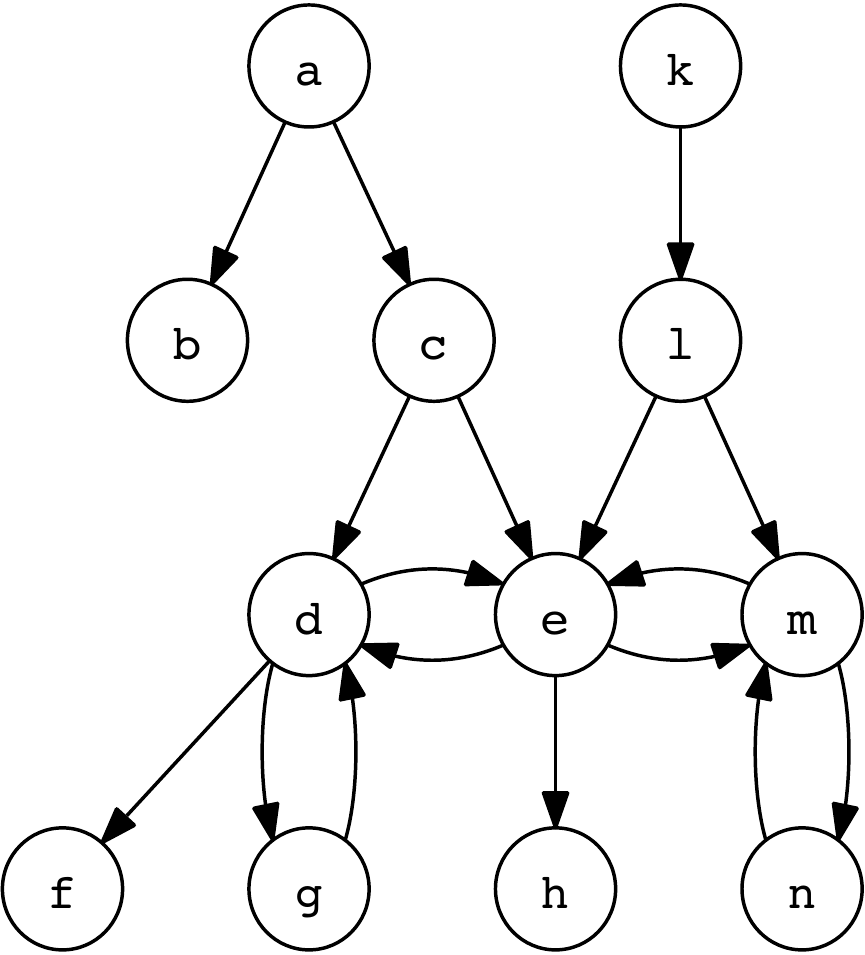}\label{fig-unsolved}}
\hfill\hfill\hfill \subfloat[The solved game reveals the answer: move
$\mathtt{e}{\to} \mathtt{h}$ is winning; the moves
$\mathtt{e}{\to}\mathtt{d}$ and $\mathtt{e}{\to}\mathtt{m}$ are not.]{
  \includegraphics[width=.37\columnwidth]{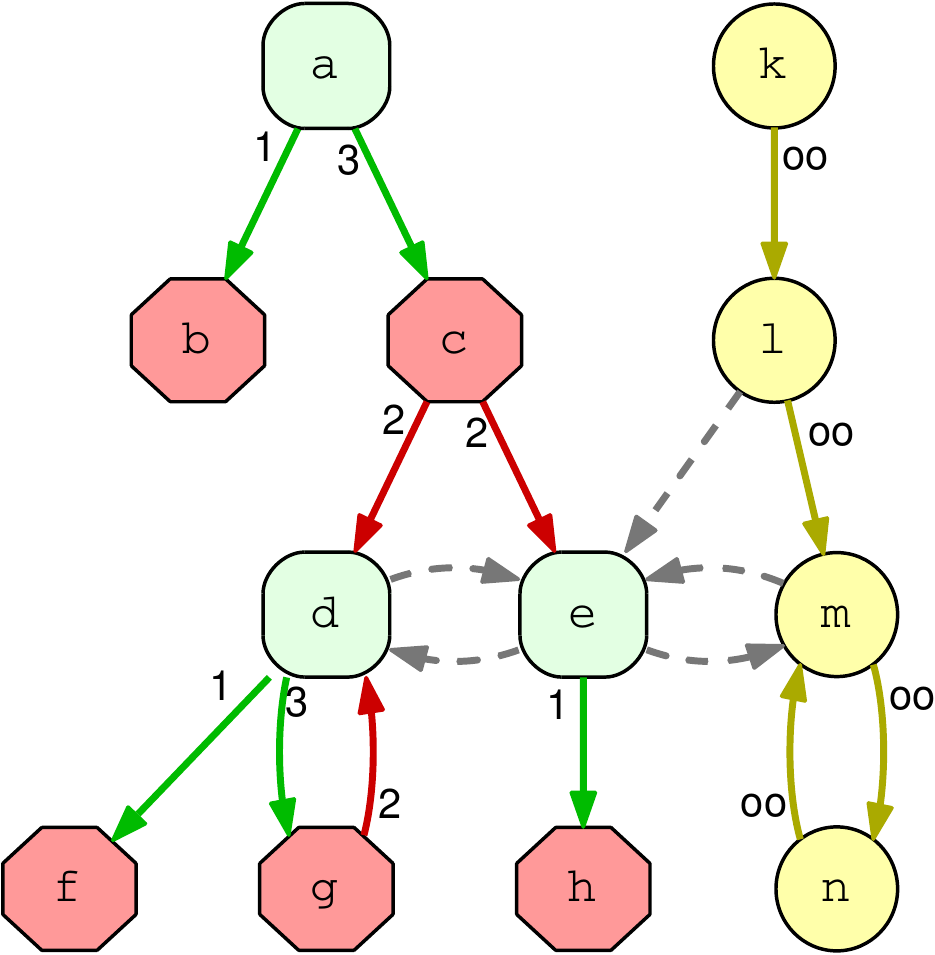}\label{fig-solved}}\hfill~
\caption{\small Position values in $G$
  (left) %
  are revealed by the solved game $\gsol=(V,M,\glabel)$ on the right:
  positions %
  are \emph{won} (green boxes), \emph{lost} (red octagons), or
  \emph{drawn} (yellow circles). This separates ``good'' moves (solid,
  colored arcs) from ``bad'' ones (dashed, gray). The length $\ell$
  of a move $x{\stackrel{\ell}{\to}}y$ indicates how quickly one can
  force a win, or how long one can delay a loss, using that move.}
  \label{fig-unsolved-solved}
\end{figure}

\mypara{Example} Consider $G=(V,M)$ in \figref{fig-unsolved} and a
start position for player \pI, say \pos{e}.  In the play $\pi_1{=} ~
\pos e \stackrel{\pI}{\to} \pos d \stackrel{\pII}{\to} \pos f$, \pI
cannot move, so $\pi_1$ is lost (for \pI). However, in $\pi_2 {=}~
\pos e \stackrel{\pI}{\to} \pos h$, \pII cannot move, so $\pi_2$ is
won (for \pI). So from position \pos e, the best move is $\pos e
{\to} \pos h$; the other moves are ``bad'': $\pos e {\to} \pos d$ loses
(see $\pi_1$), while $\pos e {\to} \pos m$ only draws (if \pII sticks to
$\pos m {\to} \pos n$).

\mypara{The Value of a Position: Playing Optimally} To determine the
true {value} of $v\in V$, we are not interested in plays with bad
moves, but consider instead plays where the opponents play optimally,
or at least ``good enough'' so that the best possible outcome is
guaranteed. Hence we ask: can \pI force a win from $v\in V$ (no matter
what \pII does), or can \pII force \pI to lose from $v$?  If neither
player can force a win, $v$ is a \emph{draw} and both players can
avoid losing by forcing an infinite play. This is formalized using
strategies.

A (pure) \emph{strategy} is a partial function $S:V\to V$ with
$S\subseteq M$.  It prescribes which of the available moves a player
will choose in a position $v$.\footnote{In our games, the same
  positions can be revisited many times. Accordingly, strategies are
  based on the current position $v$ only and do not take into account
  how one arrived at $v$.}  We define $v_0$ to be \emph{won for player
  \pI} in (at most) $n$ moves, if there is a strategy \stratI for \pI,
such that for all strategies \stratII of \pII, there is a number
$j=2k+1 \leq n$ such that
\begin{math}
  v_{j} = \stratI \circ (\stratII \circ \stratI)^k(v_0)
\end{math}
is defined, but $\stratII(v_{j})$ is not: \pII cannot move. In this
case, \stratI is a \emph{winning strategy} for \pI at $v_0$.
Conversely, $v_0$ is \emph{won for player \pII}  in (at most) $n$ moves, if
there is a strategy \stratII, such that for all strategies \stratI, 
there is a number $j=2k\leq n$  such that 
\begin{math}
 v_{j}= (\stratII \circ \stratI)^k(v_0)
\end{math}
is defined, but $\stratI(v_{j})$ is not: \pI cannot move.  With this,
we say the \emph{value} of $v_0$ is \emph{won} (\emph{lost}) if it is won for
player \pI (player \pII). If $v_0$ is neither won nor lost, its value
is \emph{drawn}, so neither \pI nor \pII can force a win from $v_0$,
but both can avoid losing via an infinite play.

\subsection{Solving Games: Labeling Nodes (Positions)}
Let $G=(V,M)$ be the game in Figure~\ref{fig-unsolved}. How
can we \emph{solve} $G$, i.e., determine whether the value of $x\in V$
is won, lost, or drawn? We represent the value of $x$ using a node
labeling $\gamma: V\to \{\won, \lost, \drawn\}$ and write $\gsol =
(V,M, \gamma)$ to denote a solved game.  

The following \datalogneg query, consisting of a single rule, solves
games:
\begin{equation}
\pos{win}(X) \la \pos{move}(X, Y), \neg \pos{win}(Y) \tag{$Q_G$}   
\end{equation}
$Q_G$ says that position $x$ is won in $G$ if there is a move to
position $y$, where $y$ is not won.  For {non-stratified}
\datalogneg programs like $Q_G$ (having recursion through negation), the
three-valued \emph{well-founded model} $\WFS$ \cite{van1991well}
provides the desired answer:
\begin{Proposition}[$Q_G$ Solves Games] 
Let $P \defeq (Q_G \cup \pos{move})$ be the \datalogneg query $Q_G$ plus
finitely many ``\pos{move}'' facts, representing a game $G=(V,M)$.
For all $x\in V$:
\begin{displaymath}
\WFS_{P}(\,\pos{win}(x)\,) = \left\{
    \begin{array}{@{}c@{}}
      \true{}\\
      \false{}\\
      \undef{}
    \end{array} \right\}
~~  \Leftrightarrow ~~
  \gamma(x) = 
  \left\{
    \begin{array}{@{}c@{}}
      \wonPos{\won}\\
      \lostPos{\lost}\\
      \drawnPos{\drawn}
    \end{array} \right\} ~. 
\end{displaymath}
\end{Proposition}
When implemented via an alternating fixpoint \cite{van1993alternating},
one obtains an increasing sequence of underestimates $U_1\subseteq U_2
\subseteq \dots$
converging to 
 the true atoms $U^\omega$
from below, and a decreasing sequence of
overestimates 
$O_1 \supseteq O_2 \supseteq \dots$ 
converging to $O^\omega$, the union of true or undefined atoms from
above. Any remaining atoms in the ``gap'' have the third truth-value
(\undef).  For the game query $Q_G$ above, $U^\omega$ contains the won
positions $V^\won$; the ``gap'' (if any) $O^\omega \setminus U^\omega$
contains the drawn positions $V^\drawn$; and the atoms in the
complement of $O^\omega$ (i.e., which are neither true nor undefined)
are the lost positions $V^\lost$. 

To solve $G$ directly, consider, e.g., the three moves $\pos
e{\to}\pos d$, $\pos e{\to} \pos h$, and $\pos e{\to} \pos m$ in
\figref{fig-unsolved}. The move $\pos e{\to} \pos h$ is
clearly winning, as it forces the opponent into a sink. However, the
status of the moves $\pos e {\to} \pos d$ and $\pos e {\to} \pos m$ is
unclear unless the game has been solved.
\figref{fig-solved} depicts the solved game \gsol. The set
of positions is a disjoint union $V = V^\won \discup V^\lost \discup
V^\drawn$.

To obtain \gsol, proceed as follows:
First, find all \emph{sinks} $x$, i.e., nodes for which the set of
\emph{followers} $\Flr(x) = \{ y \mid (x,y) \in M\}$ is empty.  
These positions are immediately lost and colored red: 
$V^\lost_0 = \{ x\in V \mid \Flr(x) = \emptyset\}$. 
In our example, $V^\lost_0 = \{ \pos b, \pos f, \pos h\}$.
We then find all nodes $x$ for which there is \emph{some} $y$ with
$(x,y)\in M$ such that $y\in V^\lost_0$. These positions are won and
colored green; here: $V^\won_1 = \{ \pos a, \pos d, \pos e \}$.
We then find the unlabeled nodes $x$ for which \emph{all} followers
$y\in \Flr(x)$ are already won (i.e., colored green). Since the player
moving from that position can only move to a position that is won for
the opponent, those $x$ are also \emph{lost} and added to $V^\lost_2$.
In our example $V^\lost_2 = \{ \pos c, \pos g\}$.  We now iterate the
above steps until there is no more change.
One can show that $V_1^\won\subseteq V_3^\won \subseteq V_5^\won
\cdots$ converges to the won positions $V^\won$, whereas $V_0^\lost
\subseteq V_2^\lost \subseteq V_4^\lost \cdots$ converges to the lost
positions $V^\lost$; the drawn positions are $V^\drawn \defeq
V\setminus (V^\won\cup V^\lost)$.

\begin{algorithm}[t]
  \SetKwData{Left}{left}\SetKwData{This}{this}\SetKwData{Up}{up}
  \SetKwFunction{Union}{Union}\SetKwFunction{FindCompress}{FindCompress}
  \SetKwInOut{Input}{Input}\SetKwInOut{Output}{Output}
  \BlankLine
  $V^\won \defeq \emptyset$  \tcp*[r]{Initially we don't know any won positions}
  $V^\lost \defeq \{ x\in V \mid \Flr(x) = \emptyset \}$ \tcp*[r]{$\dots$ but all sinks are lost $\dots$}
  $\length(x) \defeq 0$ for all $x\in V^\lost$  \tcp*[r]{$\dots$ immediately: their length is 0.}
  \Repeat{$V^\won$ and $V^\lost$ change no more}{
    \For{$x \in V \setminus (V^\won \cup V^\lost$)  
    }   
    {  %
      $F^\lost \defeq \Flr(x) \cap V^\lost$; ~~
      $F^\won \defeq \Flr(x) \cap V^\won$ \;
      \If{$F^\lost \neq \emptyset$}    
      {$V^\won \defeq V^\won \cup \{x\}$ \tcp*[r]{\textbf{\emph{some}} $y\in\Flr(x)$ is  lost, so $x$ is \textbf{\emph{won}}} 
        $\length(x) \defeq 1 + \min \{ \length(y) \mid y \in F^\lost \}$ \tcp*[r]{shortest win}}
      \If{$\Flr(x) = F^\won$}  
      {$V^\lost \defeq V^\lost \cup \{x\}$ \tcp*[r]{\textbf{\emph{all}} $y\in\Flr(x)$ are won, so $x$ is \textbf{\emph{lost}}} 
        $\length(x) \defeq 1 + \max \{ \length(y) \mid y \in F^\won \}$ \tcp*[r]{longest delay}}
    }
  } 
  
 $V^\drawn \defeq V \setminus (V^\won \cup V^\lost)$ \tcp*[r]{remaining positions are now draws}

 $\length(x) \defeq \infty$ for all $x\in V^\drawn$ \tcp*[r]{$\dots$ and can be delayed forever}

 $\glabel(x) \defeq  \won / \lost  /  \drawn$ for all $x \in  V^\won /  V^\lost /  V^\drawn$, respectively.
  \caption{Solve game $\gsol = (V,M,\glabel)$}
  \label{algo_solve_game}
\end{algorithm}

Algorithm\,\ref{algo_solve_game} depicts the details of a simple,
round-based approach to solve games.  In it, we also compute the
\emph{length} of a position, which adds further information to a
solved game \gsol, i.e., how quickly one can win (starting from green
nodes), or how long one can delay losing (starting from red nodes). In
\figref{fig-unsolved-solved}, the (delay) length of \pos f is 0, since
\pos f is a sink and no move is possible. In contrast, the (win)
length of \pos d is 1: the next player moving wins by moving to \pos
f. For \pos g, the (delay) length is 2, since the player can move to
\pos d, but the opponent can then move to \pos f. So \pos g is lost in
2 moves.

\mypara{Remark} As described, Algorithm\,\ref{algo_solve_game}
proceeds in \emph{rounds} to determine the value of positions, \ie in
each round $i$, \emph{all} newly won positions, and \emph{all} newly
lost positions are determined. This could be used, e.g., to simplify
the computation of the length of a position ($\length(x)$ can be derived
from the first round in which the value of $x$ becomes known).  On the
other hand, this is not strictly necessary: one can replace the
\textbf{for}-loop ranging over all unlabeled nodes by a
non-deterministic \textbf{pick} of any unlabeled node. As long as we
pick nodes in a fair manner, the non-deterministic version will also
converge to the correct result, while allowing more flexibility during
evaluation \cite{Zinn2012WC}.

\subsection{Game Provenance: Labeling Edges (Moves)}\label{sec-game-provenance}

\newcommand{\na}{\textcolor{gray}{\emph{n/a}}}
\begin{figure}[t]
  \centering
  \renewcommand{\arraystretch}{1.5}
~\hfill
  {\small\begin{tabular}[b]{r|c|c|c}
      & \wonPos{$y$ won}  (\won) &  \drawnPos{$y$ drawn} (\drawn)&
       \lostPos{$y$ lost} (\lost)\\
      \hline
      \wonPos{$x$ won} (\won) & \textcolor{gray}{\emph{{bad}}} & \textcolor{gray}{\emph{{bad}}} & 
      \textcolor{DarkGreen}{\textbf{\green: \emph{winning}}} \\  
      \hline
       \drawnPos{$x$ drawn} (\drawn) & \textcolor{gray}{\emph{{bad}}} & \textcolor{DarkYellow}{\textbf{\yellow: \emph{drawing}}}  &\na\\  
      \hline
       \lostPos{$x$ lost} (\lost) &
      \textcolor{DarkRed}{\textbf{\red: \emph{delaying}}}  &\na & \na \\  
    \end{tabular}}
\hfill\hfill\hfill\hfill
  \includegraphics[width=.38\columnwidth]{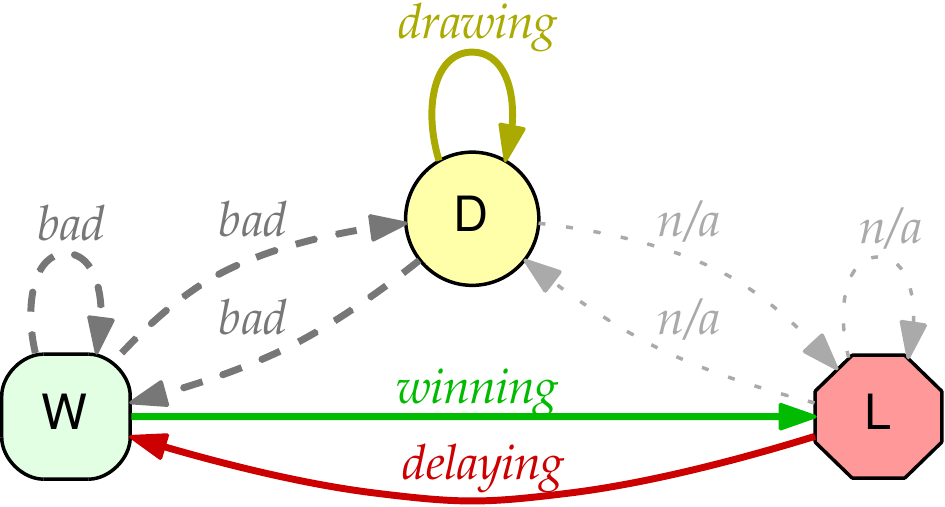}
\hfill~
  \caption{\small Depending on node labels, moves $x\to y$ are 
    either \emph{winning} (or \emph{green}) ($\won \stackrel{\green}{\leadsto} \lost$),
    \emph{delaying} (or \emph{red}) ($\lost \stackrel{\red}{\leadsto} \won$), or
    \emph{drawing} (or \emph{yellow}) ($\drawn\stackrel{\yellow}{\leadsto}\drawn$). All
    other moves are either \emph{bad} (allowing the opponent to
    improve the outcome), or non-existent (\na): e.g., if $x$ is lost,
    then there are only delaying moves (i.e., ending in won positions
    $y$ for the opponent).}
\label{fig-prov-edges}
\end{figure}
\renewcommand{\arraystretch}{1}

We return to our original question: why is $x\in V$ {won}, {lost}, or
{drawn}? We would like to define a suitable notion of \emph{game
  provenance} $\gprov(x)$ that is similar in spirit to the
how-provenance devised for positive queries
\cite{green2007provenance}, but that works for games and explains the
value (won, lost, or drawn) of $x$.
Some desiderata of game provenance are immediate: 
First, only nodes \emph{reachable} from $x$ can influence the outcome
at $x$, i.e., only nodes and edges in the transitive closure
$\Flr^+(x)$.  Thus, one expects $\gprov(x)$ to depend only on
$\Flr^+(x)$. In addition, one expects the value $\gamma(x)$ of
position $x$ to be independent of ``bad moves'', i.e., which give the
opponent a better outcome than necessary.  We use a partial edge-labeling
function $\lambda$ to distinguish different types of moves.
\begin{Definition}[Edge Labels]
  Let $\gsol=(V,M,\gamma)$ be a solved game.  The edge-labeling
  $\lambda: V\times V \to \{ \green, \red, \yellow\}$ defines a color
  for a subset of edges from $M$ as shown in \figref{fig-prov-edges}.
\end{Definition}
In \Figref{fig-prov-edges} we use $\gamma(x)$ and $\gamma(y)$, i.e.,
node labels \won, \drawn, and \lost of moves $(x,y)\in M$ to derive an
appropriate edge label.  This allows us to distinguish
provenance-relevant (``good'') moves (\emph{winning}, \emph{drawing},
or \emph{delaying}), from irrelevant (\emph{bad}) moves. The latter
are excluded from game provenance:

\begin{Definition}[Game Provenance] Let $\gsol=(V,M,\glabel)$ be a
  solved game. The \emph{game provenance} $\gprov ({=}\gprov_G)$ is
  the $\lambda$-colored subgraph of \gsol. For $x\in V$, we define
  $\gprov(x)$ as the subgraph of \gprov, reachable via $\lambda$ edges.
\end{Definition}
Consider the solved game on the right in
\figref{fig-unsolved-solved}. Since bad (dashed) edges are excluded,
the game provenance consists of two disconnected subgraphs: (i) The
bipartite ``red-green'' subgraph, which is draw-free, i.e., every
position is either won or lost, and (ii) the ``yellow'' subgraph,
representing the drawn positions.

The figure also reveals that solved games \gsol and thus game provenance
\gprov have a nice, regular structure. The following is immediate
from the underlying game-theoretic semantics of $G$.

\begin{Theorem}[Provenance Structure] Let $\gsol = (G, M, \gamma)$
  be a solved game, \gprov its edge-labeled provenance graph. The
  {game provenance} $\gprov$ has a regular structure:
\begin{equation*}
  \GHow(x) = \left\{
    \begin{array}{ll}
      M_{\green.(\red.\green)^*}(x) & \mbox{; if $x$ is won}\\
      M_{(\red.\green)^*}(x) & \mbox{; if $x$ is lost} \\
      M_{\yellow^+}(x) &  \mbox{; if $x$ is drawn}
    \end{array} \right.  %
\end{equation*}
\end{Theorem}
Here, for a regular expression $R$, and a node $x\in V$, the
expression $M_R(x)$ denotes a subset of labeled edges of $M$, i.e.,
for which there is a path $\pi$ in \gprov whose labels match the
expression $R$. As we shall see below, for positive queries, the
bipartite structure of won and lost nodes nicely corresponds to the
structure of provenance polynomials \cite{grigoris-tj-simgodrec-2012}.

\section{Provenance Games}
\label{sec-provenance-games}

The game semantics
(avoiding bad moves) yields a natural model of provenance. We now
apply this notion to queries expressed using non-recursive \datalogneg
rules. Any first-order query $\varphi(\bar x)$ on input database \D
can be expressed as a non-recursive \datalogneg program $Q_\varphi$
with a distinguished relation $\pos{ans}\in\idb{Q_\varphi}$
\footnote{The arity of \pos{ans} matches that of $\varphi(\bar
  x)$.} 
such that evaluating $Q_\varphi$ with input $D$ under the stratified
semantics\footnote{which coincides with the well-founded semantics on
  non-recursive \datalogneg} agrees with the result of $\varphi(\bar
x)$.
In the following we use $\Q(\D)$ to denote the
result of evaluating $\Q$ on input $\D$.

\subsection{Query Evaluation Games}

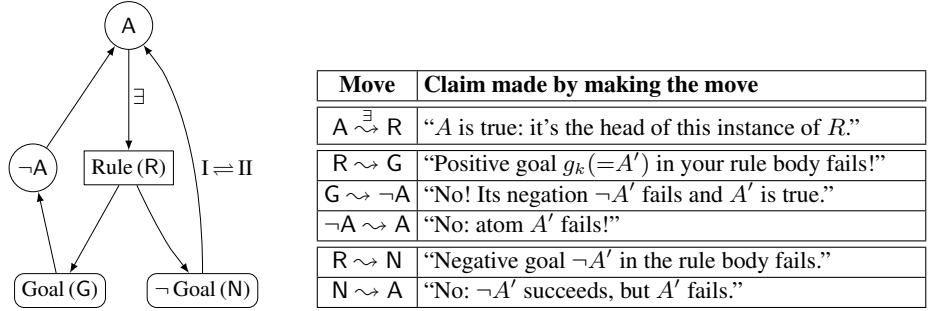
\begin{figure}[t]
  \centering
\resizebox{0.28\textwidth}{!}{\mbox{\begin{tikzpicture}[>=latex,line join=bevel,]
\node (A) at (50bp,117bp) [draw,circle] {$\,\mathsf{A}\,$};
  \node (R) at (50bp,59bp) [draw,rectangle] {Rule\,($\mathsf{R}$)};
  \node (-A) at (11bp,59bp) [draw,circle] {$\!\neg \mathsf{A}\!$};
  \node (G2) at (80bp,9bp) [draw,rectangle,rounded corners] {$\neg$\,Goal\,($\mathsf{N}$)};
  \node (G1) at (22bp,9bp) [draw,rectangle,rounded corners] {Goal\,($\mathsf{G}$)};
  \draw [->] (G1) ..controls (18.838bp,23.373bp) and (17.171bp,30.949bp)  .. (-A);
  \draw [->] (R) ..controls (41.469bp,43.766bp) and (36.347bp,34.62bp)  .. (G1);
  \definecolor{strokecol}{rgb}{0.0,0.0,0.0};
  \pgfsetstrokecolor{strokecol}
  \draw (46.5bp,33bp) node { };
  \draw [->] (A) ..controls (50bp,97.86bp) and (50bp,86.981bp)  .. (R);
  \draw (56.5bp,88bp) node {$\!\!\!\exists$};
  \draw [->] (R) ..controls (57.146bp,43.566bp) and (61.968bp,33.951bp)  .. (67bp,26bp) .. controls (67.055bp,25.914bp) and (67.11bp,25.827bp)  .. (G2);
  \draw (73.5bp,33bp) node { };
  \draw [->] (G2) ..controls (80.1bp,34.25bp) and (78.234bp,71.252bp)  .. (64bp,98bp) .. controls (63.639bp,98.678bp) and (63.256bp,99.359bp)  .. (A);
  \draw (90bp,59bp) node {I\,$\rightleftharpoons$\,II};
  \draw [->] (-A) ..controls (22.956bp,76.78bp) and (31.201bp,89.043bp)  .. (A);
\end{tikzpicture}}}
\hfill
{\footnotesize\begin{tabular}[b]{|@{~}c@{~}|@{~}p{0.54\textwidth}@{~}|}
\hline
\textbf{Move} & \textbf{Claim made by making the move}\\
\hline
\hline
  $\mathsf{A}\stackrel{\exists}{\leadsto} \mathsf{R}$ & ``$A$ is true: it's the head of this instance of ${R}$.''\\
\hline
\hline
  $\mathsf{R}\leadsto \mathsf{G}$ & ``Positive goal $g_k ({=}A')$ in your rule body fails!''\\
\hline
  $\mathsf{G}\leadsto\neg \mathsf{A}$ & ``No! Its negation $\neg A'$
  fails and $A'$ is true.''\\
\hline
 $\neg \mathsf{A}\leadsto \mathsf{A}$ & ``No:  atom $A'$ fails!''\\
\hline
\hline
  $\mathsf{R}\leadsto \mathsf{N}$ & ``Negative goal $\neg A'$ in the rule body fails.''\\
\hline
  $\mathsf{N}\leadsto \mathsf{A}$ & ``No: $\neg A'$ succeeds, but $A'$ fails.''\\
\hline
\end{tabular}}
\caption{\small Move types of the query evaluation game (left) and
  implicit claims made (right). Moving along an edge, a player aims to
  verify a claim, thereby refuting the opponent. Initially, player \pI
  is a verifier, trying to prove $A$, while \pII tries to spoil this
  attempt and refute it. Roles are swapped (\pI $\rightleftharpoons$
  \pII) when moving through a negated goal ($\mathsf{R}{\leadsto}\mathsf{N}{\leadsto}\mathsf{A}$).}
  \label{fig-game-overview}
\end{figure}

Query evaluation of $\Q(\D)$ can be seen as a game between players \pI
and \pII who argue whether an atom $A\in\Q(\D)$. 
The argumentation structure is stylized in \figref{fig-game-overview}. There are
three classes of positions in the game as shown on the left of
\Figref{fig-game-overview}:
\begin{itemize} 
  \item Relation nodes---depicted as circles,
  \item Rule nodes---depicted as rectangles, and
  \item Goal nodes---depicted as rectangles with rounded corners.
\end{itemize}
Both relation nodes and goal nodes can be positive or negative. 

Usually, an evaluation game starts with I claiming that a ground atom
$A(x)$ is true. That is she starts the game in a relation node for $A$. 
To substantiate her 
claim she moves to a rule that has $A$ as a head atom and specifies
constants for the remaining
existentially quantified variables in the body of the rule. Now, II tries to reject
the validity of the rule by selecting a goal atom (\eg $B$) in its body that he thinks is not
satisfied (\eg II moves to the goal node for $B$).
I then moves to a negated relation node for this goal
(eg, a node $\neg B$), claiming the goal is
true because its negation is false. From here, II moves to the relation node
$B$, questioning I's claim that $B$ is true. The game then
continues in the same way. Note that the graph on the left of in \figref{fig-game-overview} is a schema-level
description. When one cycle
(relation$\leadsto$rule$\leadsto$goal$\leadsto$$\neg$relation$\leadsto$relation) is complete, the actual fact that 
is argued about has changed (\eg from $A$ to $B$). If II selects a negated goal (\eg $\neg C$) in the body of a rule then player I moves directly
from the negated goal node to the relation node for $C$. This essentially switches the roles of I
and II since now player II has to argue for a relation node $C$.

\begin{figure}
  \centering
  \subfloat[\Pabc as a game diagram]{
    \mbox{\resizebox{0.42\columnwidth}{!}{\begin{tikzpicture}[>=latex,line join=bevel,]
\node (a) at (88bp,226bp) [draw,ellipse, fill=white, draw] {$A(X)$};
  \node (c) at (146bp,60bp) [draw,ellipse, fill=white, draw] {$C(X)$};
  \node (b) at (60bp,60bp) [draw,ellipse, fill=white, draw] {$B(X,Y)$};
  \node (r1r) at (88bp,177bp) [draw,rectangle] {${r_1}(X,Y)$};
  \node (r1b1) at (62bp,141bp) [draw=black,rectangle, fill=white, draw,rounded corners] {${g^1_1}(X,Y)$};
  \node (r1b2) at (107bp,141bp) [draw=black,rectangle, fill=white, draw,rounded corners] {${g^2_1}(Y)$};
  \node (rb) at (60bp,9bp) [draw=black,rectangle, fill=white, draw] {${r_B}(X,Y)$};
  \node (rc) at (146bp,9bp) [draw=black,rectangle, fill=white, draw] {${r_C}(X)$};
  \node (n_a) at (26bp,226bp) [draw=black,ellipse] {${\neg{}A}(X)$};
  \node (n_b) at (60bp,102bp) [draw=black,ellipse] {${\neg{}B}(X,Y)$};
  \node (n_c) at (172bp,102bp) [draw=black,ellipse] {${\neg{}C}(X)$};
  \draw [->] (r1r) ..controls (94.049bp,165.54bp) and (95.708bp,162.39bp)  .. (r1b2);
  \draw [->] (b) ..controls (60bp,41.951bp) and (60bp,34.62bp)  .. (rb);
  \definecolor{strokecol}{rgb}{0.0,0.0,0.0};
  \pgfsetstrokecolor{strokecol}
  \draw (80.5bp,33bp) node {$B(X,Y)$};
  \draw [->] (n_b) ..controls (60bp,87.298bp) and (60bp,84.8bp)  .. (b);
  \draw [->] (r1b1) ..controls (61.417bp,129.63bp) and (61.283bp,127.03bp)  .. (n_b);
  \draw [->] (n_a) ..controls (53.3bp,226bp) and (54.443bp,226bp)  .. (a);
  \draw [->] (r1r) ..controls (79.536bp,165.28bp) and (77.023bp,161.8bp)  .. (r1b1);
  \draw [->] (c) ..controls (146bp,41.951bp) and (146bp,34.62bp)  .. (rc);
  \draw (165.5bp,33bp) node {$C(X)\quad$};
  \draw [->] (r1b2) ..controls (105.71bp,121.12bp) and (106.15bp,103.32bp)  .. (113bp,90bp) .. controls (114.99bp,86.127bp) and (117.7bp,82.508bp)  .. (c);
  \draw (127.5bp,102bp) node {X:=Y};
  \draw [->] (n_c) ..controls (162.82bp,87.176bp) and (160.85bp,83.988bp)  .. (c);
  \draw [->] (a) ..controls (88bp,208.43bp) and (88bp,201.86bp)  .. (r1r);
  \draw (96.5bp,200bp) node {$\exists$Y};
\end{tikzpicture}}}
    \label{abcgamediagram}
  }
  \subfloat[Rules defining the moves for \Pabc]
 {
    \parbox[b][][t]{.5\textwidth}{
    \begin{displaymath} 
      \arraycolsep=1.5pt
      \begin{array}{lcl}
        \textnormal{\textbf{Atoms} $A$,$B$, \textbf{and} $C$} \\
        \mrel{M}(\mskol{\neg{}A}{X}, \mskol{A}{X}) &\la& \Domain(X).\\
        \mrel{M}(\mskol{\neg{}B}{X,Y},\mskol{B}{X,Y}) &\la&  \Domain(X),\Domain(Y). \\
        \mrel{M}(\mskol{\neg{}C}{X},\mskol{C}{X}) &\la& \Domain(X). \\ \\

        \textnormal{\textbf{IDB} $A$ \textbf{via rule} $r_1$} \\
        \mrel{M}(\mskol{A}{X},\mskol{r_1}{X,Y}) &\la& \Domain(X), \Domain(Y).\\
        \mrel{M}(\mskol{r_1}{X,Y}, \mskol{g_1^1}{X,Y}) &\la& \Domain(X), \Domain(Y).\\
        \mrel{M}(\mskol{r_1}{X,Y}, \mskol{g_1^2}{Y}) &\la&\Domain(X), \Domain(Y).\\
        \mrel{M}(\mskol{g_1^1}{X,Y} ,\mskol{\neg{}B}{X,Y}) &\la& \Domain(X), \Domain(Y).\\
        \mrel{M}(\mskol{g_1^2}{X},\mskol{C}{X}) &\la& \Domain(X). \\ \\
        \textnormal{\textbf{EDB} $B$ \textbf{and} $C$} \\
        \mrel{M}(\mskol{B}{X,Y},\mskol{r_B}{X,Y}) &\la& \mathtt{B}(X,Y).\\ 
        \mrel{M}(\mskol{C}{X},\mskol{r_C}X) &\la& \mathtt{C}(X).
        \end{array}
      \end{displaymath}
    }
    \label{fig-reduction}
  }
  \\

  \subfloat[Instantiated game $G_\PabcD$ for $\D= \{ B(a,b), B(b,a), C(a)\}$]{
     \mbox{\resizebox{.7\columnwidth}{!}{\begin{tikzpicture}[>=latex,line join=bevel,]
\node (n_c_a) at (163bp,157bp) [draw=black,ellipse, draw] {$\neg{}C(a)$};
  \node (n_c_b) at (163bp,85bp) [draw=black,ellipse, draw] {$\neg{}C(b)$};
  \node (n_b_aa) at (238bp,40bp) [draw=black,ellipse, draw] {$\neg{}B(a,a)$};
  \node (n_b_ab) at (238bp,12bp) [draw=black,ellipse, draw] {$\neg{}B(a,b)$};
  \node (rBr_ba) at (380bp,182bp) [draw,rectangle,  draw] {$r_B(b,a)$};
  \node (r1r_ba) at (100bp,142bp) [draw,rectangle,  draw] {$r_1(b,a)$};
  \node (n_a_b) at (24bp,145bp) [draw=black,ellipse,  draw] {$\neg{}A(b)$};
  \node (n_a_a) at (24bp,29bp) [draw=black,ellipse,  draw] {$\neg{}A(a)$};
  \node (r1b1_aa) at (163bp,38bp) [draw=black,rectangle,  draw] {$g_1^1(a,a)$};
  \node (b_ab) at (315bp,12bp) [draw,ellipse, draw] {$B(a,b)$};
  \node (b_aa) at (315bp,40bp) [draw,ellipse, draw] {$B(a,a)$};
  \node (c_a) at (238bp,147bp) [draw,ellipse, draw] {$C(a)$};
  \node (r1b2_ca) at (163bp,132bp) [draw=black,rectangle,  draw] {$g_1^2(a)$};
  \node (r1b2_cb) at (163bp,60bp) [draw=black,rectangle,  draw] {$g_1^2(b)$};
  \node (c_b) at (238bp,75bp) [draw,ellipse, draw] {$C(b)$};
  \node (n_b_ba) at (238bp,182bp) [draw=black,ellipse, draw] {$\neg{}B(b,a)$};
  \node (n_b_bb) at (238bp,110bp) [draw=black,ellipse, draw] {$\neg{}B(b,b)$};
  \node (rCr_a) at (315bp,147bp) [draw,rectangle,  draw] {$r_C(a)$};
  \node (a_b) at (24bp,104bp) [draw,ellipse, draw] {$A(b)$};
  \node (a_a) at (24bp,70bp) [draw,ellipse, draw] {$A(a)$};
  \node (r1r_ab) at (100bp,34bp) [draw,rectangle,  draw] {$r_1(a,b)$};
  \node (r1r_aa) at (100bp,70bp) [draw,rectangle,  draw] {$r_1(a,a)$};
  \node (r1b1_ab) at (163bp,13bp) [draw=black,rectangle,  draw] {$g_1^1(a,b)$};
  \node (rBr_ab) at (380bp,12bp) [draw,rectangle,  draw] {$r_B(a,b)$};
  \node (r1r_bb) at (100bp,104bp) [draw,rectangle,  draw] {$r_1(b,b)$};
  \node (r1b1_bb) at (163bp,110bp) [draw=black,rectangle,  draw] {$g_1^1(b,b)$};
  \node (r1b1_ba) at (163bp,182bp) [draw=black,rectangle,  draw] {$g_1^1(b,a)$};
  \node (b_bb) at (315bp,110bp) [draw,ellipse, draw] {$B(b,b)$};
  \node (b_ba) at (315bp,182bp) [draw,ellipse, draw] {$B(b,a)$};
  \draw [->] (r1b1_aa) ..controls (186.95bp,38.629bp) and (191.51bp,38.754bp)  .. (n_b_aa);
  \draw [->] (r1r_ba) ..controls (117.57bp,154.88bp) and (125.1bp,160.45bp)  .. (r1b1_ba);
  \draw [->] (a_a) ..controls (51.736bp,70bp) and (61.352bp,70bp)  .. (r1r_aa);
  \definecolor{strokecol}{rgb}{0.0,0.0,0.0};
  \pgfsetstrokecolor{strokecol}
  \draw (64bp,75.5bp) node {$\exists$a};
  \draw [->] (n_a_a) ..controls (24bp,43.229bp) and (24bp,45.528bp)  .. (a_a);
  \draw (14.5bp,49.5bp) node { };
  \draw [->] (n_b_ba) ..controls (272bp,182bp) and (275.44bp,182bp)  .. (b_ba);
  \draw [->] (n_b_bb) ..controls (271.86bp,110bp) and (275.62bp,110bp)  .. (b_bb);
  \draw [->] (r1b1_ab) ..controls (186.68bp,12.689bp) and (191.85bp,12.619bp)  .. (n_b_ab);
  \draw [->] (n_b_aa) ..controls (272.3bp,40bp) and (275.15bp,40bp)  .. (b_aa);
  \draw [->] (n_a_b) ..controls (24bp,130.71bp) and (24bp,128.41bp)  .. (a_b);
  \draw (14.5bp,124.5bp) node { };
  \draw [->] (b_ba) ..controls (343.96bp,182bp) and (346.72bp,182bp)  .. (rBr_ba);
  \draw [->] (r1r_ba) ..controls (125.15bp,138.04bp) and (131.71bp,136.97bp)  .. (r1b2_ca);
  \draw [->] (r1r_ab) ..controls (124.01bp,26.08bp) and (129.18bp,24.3bp)  .. (r1b1_ab);
  \draw [->] (n_b_ab) ..controls (272bp,12bp) and (275.44bp,12bp)  .. (b_ab);
  \draw [->] (r1b1_bb) ..controls (186.58bp,110bp) and (192.11bp,110bp)  .. (n_b_bb);
  \draw [->] (r1r_bb) ..controls (117.27bp,88.993bp) and (127.88bp,79.877bp)  .. (138bp,73bp) .. controls (138.39bp,72.734bp) and (138.79bp,72.468bp)  .. (r1b2_cb);
  \draw [->] (c_a) ..controls (267.07bp,147bp) and (278.82bp,147bp)  .. (rCr_a);
  \draw [->] (r1r_aa) ..controls (122.72bp,58.606bp) and (129.29bp,55.161bp)  .. (r1b1_aa);
  \draw [->] (r1b2_cb) ..controls (186.56bp,64.635bp) and (198.34bp,67.056bp)  .. (c_b);
  \draw [->] (a_b) ..controls (51.046bp,104bp) and (61.474bp,104bp)  .. (r1r_bb);
  \draw (64bp,109.5bp) node {$\exists$b};
  \draw [->] (r1b2_ca) ..controls (186.65bp,136.65bp) and (197.92bp,138.97bp)  .. (c_a);
  \draw [->] (b_ab) ..controls (343.96bp,12bp) and (346.72bp,12bp)  .. (rBr_ab);
  \draw [->] (a_a) ..controls (52.408bp,56.666bp) and (62.965bp,51.53bp)  .. (r1r_ab);
  \draw (64bp,60.5bp) node {$\exists$b};
  \draw [->] (r1r_aa) ..controls (111.84bp,82.681bp) and (116.25bp,88.045bp)  .. (120bp,93bp) .. controls (128.48bp,104.21bp) and (127.53bp,109.62bp)  .. (138bp,119bp) .. controls (138.35bp,119.32bp) and (138.71bp,119.63bp)  .. (r1b2_ca);
  \draw [->] (r1r_bb) ..controls (123.41bp,106.2bp) and (128.68bp,106.72bp)  .. (r1b1_bb);
  \draw [->] (n_c_b) ..controls (193.75bp,80.924bp) and (201.61bp,79.847bp)  .. (c_b);
  \draw [->] (n_c_a) ..controls (194.34bp,152.84bp) and (201.38bp,151.88bp)  .. (c_a);
  \draw [->] (r1r_ab) ..controls (125.38bp,44.392bp) and (132.24bp,47.314bp)  .. (r1b2_cb);
  \draw [->] (r1b1_ba) ..controls (186.68bp,182bp) and (191.85bp,182bp)  .. (n_b_ba);
  \draw [->] (a_b) ..controls (51.896bp,117.81bp) and (63.577bp,123.81bp)  .. (r1r_ba);
  \draw (64bp,133.5bp) node {$\exists$a};
\end{tikzpicture}}}
     \label{gamegraphpabc}
  }

  \subfloat[
  Solved game $\gsol_\PabcD$. %
  Lost positions are (dark) red; won positions are (light) green.
  Provenance edges (good moves) are solid; bad moves are dashed.
$A(a)$ (resp.\ $A(b)$) is true (resp.\ false), indicated by position value \won
(resp.\ \lost). 
The game provenance $\gprov(A(a))$ explains why/how $A(a)$ is true; 
$\gprov(A(b))$ explains why-not $A(b)$.
]
 {
     \mbox{\resizebox{.7\columnwidth}{!}{\begin{tikzpicture}[>=latex,line join=bevel,]
\node (n_c_a) at (163bp,157bp) [draw=black,ellipse, fill=red!40, draw] {$\neg{}C(a)$};
  \node (n_c_b) at (163bp,85bp) [draw=black,ellipse, fill=green!11, draw] {$\neg{}C(b)$};
  \node (n_b_aa) at (238bp,40bp) [draw=black,ellipse, fill=green!11, draw] {$\neg{}B(a,a)$};
  \node (n_b_ab) at (238bp,12bp) [draw=black,ellipse, fill=red!40, draw] {$\neg{}B(a,b)$};
  \node (rBr_ba) at (380bp,182bp) [draw,rectangle, fill=red!40, draw] {$r_B(b,a)$};
  \node (r1r_ba) at (100bp,142bp) [draw,rectangle, fill=green!11, draw] {$r_1(b,a)$};
  \node (n_a_b) at (24bp,145bp) [draw=black,ellipse, fill=green!11, draw] {$\neg{}A(b)$};
  \node (n_a_a) at (24bp,29bp) [draw=black,ellipse, fill=red!40, draw] {$\neg{}A(a)$};
  \node (r1b1_aa) at (163bp,38bp) [draw=black,rectangle, fill=red!40, draw] {$g_1^1(a,a)$};
  \node (b_ab) at (315bp,12bp) [draw,ellipse, fill=green!11, draw] {$B(a,b)$};
  \node (b_aa) at (315bp,40bp) [draw,ellipse, fill=red!40, draw] {$B(a,a)$};
  \node (c_a) at (238bp,147bp) [draw,ellipse, fill=green!11, draw] {$C(a)$};
  \node (r1b2_ca) at (163bp,132bp) [draw=black,rectangle, fill=red!40, draw] {$g_1^2(a)$};
  \node (r1b2_cb) at (163bp,60bp) [draw=black,rectangle, fill=green!11, draw] {$g_1^2(b)$};
  \node (c_b) at (238bp,75bp) [draw,ellipse, fill=red!40, draw] {$C(b)$};
  \node (n_b_ba) at (238bp,182bp) [draw=black,ellipse, fill=red!40, draw] {$\neg{}B(b,a)$};
  \node (n_b_bb) at (238bp,110bp) [draw=black,ellipse, fill=green!11, draw] {$\neg{}B(b,b)$};
  \node (rCr_a) at (315bp,147bp) [draw,rectangle, fill=red!40, draw] {$r_C(a)$};
  \node (a_b) at (24bp,104bp) [draw,ellipse, fill=red!40, draw] {$A(b)$};
  \node (a_a) at (24bp,70bp) [draw,ellipse, fill=green!11, draw] {$A(a)$};
  \node (r1r_ab) at (100bp,34bp) [draw,rectangle, fill=red!40, draw] {$r_1(a,b)$};
  \node (r1r_aa) at (100bp,70bp) [draw,rectangle, fill=green!11, draw] {$r_1(a,a)$};
  \node (r1b1_ab) at (163bp,13bp) [draw=black,rectangle, fill=green!11, draw] {$g_1^1(a,b)$};
  \node (rBr_ab) at (380bp,12bp) [draw,rectangle, fill=red!40, draw] {$r_B(a,b)$};
  \node (r1r_bb) at (100bp,104bp) [draw,rectangle, fill=green!11, draw] {$r_1(b,b)$};
  \node (r1b1_bb) at (163bp,110bp) [draw=black,rectangle, fill=red!40, draw] {$g_1^1(b,b)$};
  \node (r1b1_ba) at (163bp,182bp) [draw=black,rectangle, fill=green!11, draw] {$g_1^1(b,a)$};
  \node (b_bb) at (315bp,110bp) [draw,ellipse, fill=red!40, draw] {$B(b,b)$};
  \node (b_ba) at (315bp,182bp) [draw,ellipse, fill=green!11, draw] {$B(b,a)$};
  \draw [->,line width=1,red] (r1b1_aa) ..controls (186.95bp,38.629bp) and (191.51bp,38.754bp)  .. (n_b_aa);
  \draw [->,line width=1,gray,dashed] (r1r_ba) ..controls (117.57bp,154.88bp) and (125.1bp,160.45bp)  .. (r1b1_ba);
  \draw [->,line width=1,gray,dashed] (a_a) ..controls (51.736bp,70bp) and (61.352bp,70bp)  .. (r1r_aa);
  \definecolor{strokecol}{rgb}{0.0,0.0,0.0};
  \pgfsetstrokecolor{strokecol}
  \draw (64bp,75.5bp) node {$\exists$a};
  \draw [->,line width=1,red] (n_a_a) ..controls (24bp,43.229bp) and (24bp,45.528bp)  .. (a_a);
  \draw (14.5bp,49.5bp) node { };
  \draw [->,line width=1,red] (n_b_ba) ..controls (272bp,182bp) and (275.44bp,182bp)  .. (b_ba);
  \draw [->,line width=1.5,green!50!black] (n_b_bb) ..controls (271.86bp,110bp) and (275.62bp,110bp)  .. (b_bb);
  \draw [->,line width=1.5,green!50!black] (r1b1_ab) ..controls (186.68bp,12.689bp) and (191.85bp,12.619bp)  .. (n_b_ab);
  \draw [->,line width=1.5,green!50!black] (n_b_aa) ..controls (272.3bp,40bp) and (275.15bp,40bp)  .. (b_aa);
  \draw [->,line width=1.5,green!50!black] (n_a_b) ..controls (24bp,130.71bp) and (24bp,128.41bp)  .. (a_b);
  \draw (14.5bp,124.5bp) node { };
  \draw [->,line width=1,red] (b_ba) ..controls (343.96bp,182bp) and (346.72bp,182bp)  .. (rBr_ba);
  \draw [->,line width=1.5,green!50!black] (r1r_ba) ..controls (125.15bp,138.04bp) and (131.71bp,136.97bp)  .. (r1b2_ca);
  \draw [->,line width=1,red] (r1r_ab) ..controls (124.01bp,26.08bp) and (129.18bp,24.3bp)  .. (r1b1_ab);
  \draw [->,line width=1,red] (n_b_ab) ..controls (272bp,12bp) and (275.44bp,12bp)  .. (b_ab);
  \draw [->,line width=1,red] (r1b1_bb) ..controls (186.58bp,110bp) and (192.11bp,110bp)  .. (n_b_bb);
  \draw [->,line width=1,gray,dashed] (r1r_bb) ..controls (117.27bp,88.993bp) and (127.88bp,79.877bp)  .. (138bp,73bp) .. controls (138.39bp,72.734bp) and (138.79bp,72.468bp)  .. (r1b2_cb);
  \draw [->,line width=1,red] (c_a) ..controls (267.07bp,147bp) and (278.82bp,147bp)  .. (rCr_a);
  \draw [->,line width=1.5,green!50!black] (r1r_aa) ..controls (122.72bp,58.606bp) and (129.29bp,55.161bp)  .. (r1b1_aa);
  \draw [->,line width=1.5,green!50!black] (r1b2_cb) ..controls (186.56bp,64.635bp) and (198.34bp,67.056bp)  .. (c_b);
  \draw [->,line width=1,red] (a_b) ..controls (51.046bp,104bp) and (61.474bp,104bp)  .. (r1r_bb);
  \draw (64bp,109.5bp) node {$\exists$b};
  \draw [->,line width=1,red] (r1b2_ca) ..controls (186.65bp,136.65bp) and (197.92bp,138.97bp)  .. (c_a);
  \draw [->,line width=1,red] (b_ab) ..controls (343.96bp,12bp) and (346.72bp,12bp)  .. (rBr_ab);
  \draw [->,line width=1.5,green!50!black] (a_a) ..controls (52.408bp,56.666bp) and (62.965bp,51.53bp)  .. (r1r_ab);
  \draw (64bp,60.5bp) node {$\exists$b};
  \draw [->,line width=1.5,green!50!black] (r1r_aa) ..controls (111.84bp,82.681bp) and (116.25bp,88.045bp)  .. (120bp,93bp) .. controls (128.48bp,104.21bp) and (127.53bp,109.62bp)  .. (138bp,119bp) .. controls (138.35bp,119.32bp) and (138.71bp,119.63bp)  .. (r1b2_ca);
  \draw [->,line width=1.5,green!50!black] (r1r_bb) ..controls (123.41bp,106.2bp) and (128.68bp,106.72bp)  .. (r1b1_bb);
  \draw [->,line width=1.5,green!50!black] (n_c_b) ..controls (193.75bp,80.924bp) and (201.61bp,79.847bp)  .. (c_b);
  \draw [->,line width=1,red] (n_c_a) ..controls (194.34bp,152.84bp) and (201.38bp,151.88bp)  .. (c_a);
  \draw [->,line width=1,red] (r1r_ab) ..controls (125.38bp,44.392bp) and (132.24bp,47.314bp)  .. (r1b2_cb);
  \draw [->,line width=1.5,green!50!black] (r1b1_ba) ..controls (186.68bp,182bp) and (191.85bp,182bp)  .. (n_b_ba);
  \draw [->,line width=1,red] (a_b) ..controls (51.896bp,117.81bp) and (63.577bp,123.81bp)  .. (r1r_ba);
  \draw (64bp,133.5bp) node {$\exists$a};
\end{tikzpicture}}}
     \label{solvedgamegraphpabc}
  }

  \caption{\small Provenance game for the FO query $\Pabc {\defeq} ~  {\rel A(X) \la \rel
  B(X,Y), \neg \rel C(Y).}$ The well-founded
model of the rule $\pos{win}(X) \la \pos{M}(X, Y), \neg
\pos{win}(Y)$, applied to the move graph \pos M,  solves the game.
\label{pabcgraph}}
 \end{figure}
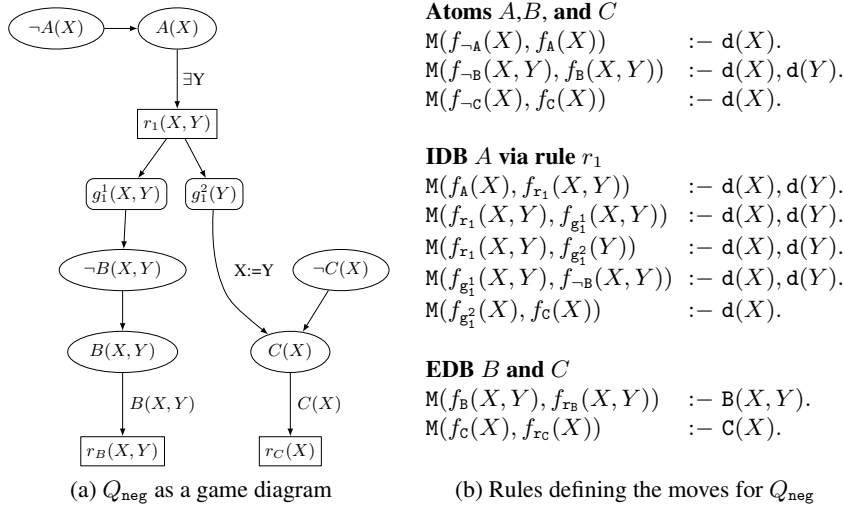
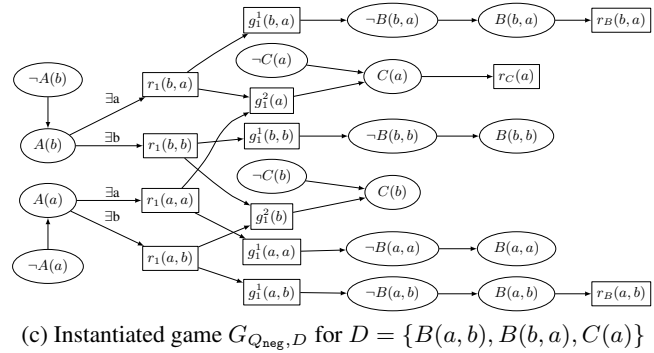
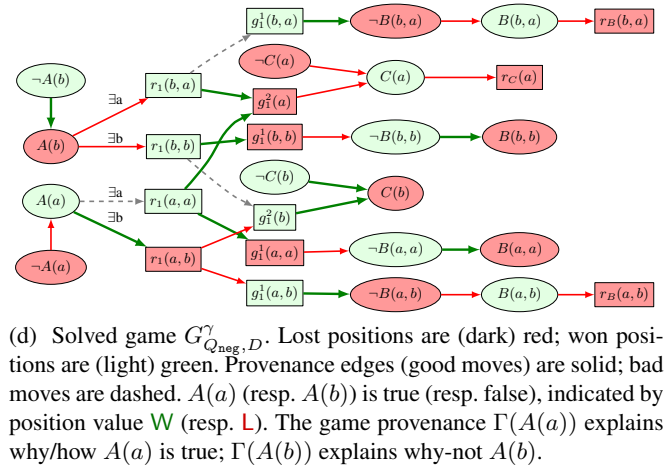

We now demonstrate the general argumentation scheme for a concrete \datalogneg
program \Pabc. The program $\Pabc$ consists of a single rule $r_1$: 
\begin{equation}
r_1: \quad \rel{A}(X)  \la  
  \underbrace{\rel{B}(X,Y)}_{\mathnormal{g_1}},\;
  \underbrace{\neg\,\rel{C}(Y)}_{\mathnormal{g_2}} 
  \tag{\Pabc}
\end{equation}
The game diagram for \Pabc is shown in \figref{abcgamediagram}. Player \pI
starts in a relation node of type $A(X)$ with a concrete instatiation $X=x$ 
to prove that $\rel A(x)\in\Q(\D)$. 
In her first move, she picks the rule $r_1$ together with
bindings for all existentially quantified variables in $r_1$, 
which is just a instatiation $y$  for $Y$ in $r_1$; 
essentially
picking a ground instance $r_1(x,y)$ such that the variable $X$ is bound to the
desired $x$. She claims the rule body is satisfied.
If this is not the case, \pII can falsify the claim by
selecting a \emph{goal} from the body, i.e., either $g_1^1(x,y)$, %
thus making a counter-claim that $\rel B(x,y)$ is false, or %
$g_1^2(y)$, claiming instead that $C(y)$ is true. \textit{Positive case,} \eg
\pII moved to $g_1^1(x,y)$.  Player \pI
will move from $g_1^1(x,y)$ to $\neg B(x,y)$, from which \pII will move to
$B(x,y)$. In this node, there is an edge for player \pI if and only if $B(x,y)
\in D$, that is if there is a trivial, bodyless rule $r_B(x,y)$ representing
this fact. Thus, \pI wins the game if $B(x,y) \in D$ and \pII wins if $B(x,y)
\not\in D$. \textit{Negative case,} \eg \pII just moved to $g_1^2(y)$. Player
\pI moves to the instatiation $\mathtt{C}(y)$ of relation node $C(X)$. 
For this move in the diagram, variables used in the goal node are explicitely 
renamed to the single variable name used in the corresponding relation node.
With this move, \pII loses and \pI wins if $C(y) \not\in D$;
\pII wins the argument if $C(y) \in D$ by moving to the trivial rule node,
forcing \pI to lose.

\mypara{Construction of Evaluation Game Graph}
We create a game in which the constants are also encoded within the
game positions. In \figref{fig-reduction}, we provide Datalog rules that
define the move relation $\mathtt{M}$ of the evaluation game $G_{\Pabc,D}$ for \Pabc with an input database $D$.
Here, $\Domain$ is a relation that contains the active domain of \Pabc and $D$.

For each ground atom, we create a postive and a negative relation node. 
We use Skolem functions to create ``node identifiers''. E.g., for a ground atom
$\mathtt{S(a_1,\dots,a_n)}$ we use $f_S(\mathtt{a_1,\dots,a_n})$ for its positive relation nodes and 
$f_{\neg{}S}(\mathtt{a_1,\dots,a_n})$ for its negative relation node. The first three
rules in \figref{fig-reduction} create an edge from the negative to the
positive node.\footnote{The use of Skolems is for convenience only.
We could instead use constants and increase the arity of relations accordingly, 
or even avoid constants \cite{flum-ICDT-97,flum-TCS-00}.}

Furthermore, we create a \emph{rule node} for each rule $r_i$ in the ground program 
with a unique identifier $f_{r_i}(X_1,\dots,X_n)$ including the rule number and 
the assignments of variables found in the rule's body to constants. For simplicity, we
alphabetically order variables and provide the constants in this order. There is
an edge from the ground head atom to the ground rule node (cf.
\figref{fig-reduction} first line of middle block).
For example, the Skolem function $f_{r_1}(a,b)$ encodes the whole rule body $r1:[B(a,b),\neg{}C(b)]$.

Then, we add moves from rule node $r_i$ to its \emph{goal nodes} $g_i^j$. 
Goal nodes are identified by the rule number $i$ they occur in, their
positions $j$ within the body, and the bound constants. (cf. lines 2 and 3 of
middle block). From positive (negative) goal nodes, we move to negative (positive) relation
nodes keeping the bound constants fixed (cf. lines 4 and 5 of middle block).
Finally, for edb relations, we add an edge from the positive relation node
$R(\bar c)$ to a rule node $\mskol{r_R}{\bar c}$ iff $R(\bar c) \in D$. This
ensures that a player reaching the relation node $R(\bar c)$ wins iff $R(\bar c)
\in D$.
In \figref{gamegraphpabc} the game graph for \Pabc with input database $\D= \{
B(a,b), B(b,a), C(a)\}$ is shown. The solved game is shown in
\figref{solvedgamegraphpabc}. Here, we see that \pI has a winning strategy for
\eg $A(a)$, $B(b,a)$, and $C(a)$.

\mypara{Acyclicity of FO Games}
For FO queries, represented by non-recursive \datalogneg programs,
no relation node is reachable from itself and the resulting game graph is acyclic.
\begin{Theorem}[FO Provenance Game]
  Consider a first-order query $\varphi$ in the form of a
  non-recursive \datalogneg program $\Prog{\varphi}$ with output relation
  \rel{ans} and input database facts $\D$. 
  Let  $\gsol_\ProgD{\varphi}{\D} = (V,M,\gamma)$ be the
  solved game. Then:
  \begin{enumerate}
  \item $\gsol_\ProgD{\varphi}{\D}$ is draw-free.
  \item 
\begin{math}
  \Prog{\varphi}(\, \rel{ans}(\bar x)\, )  = \left\{
    \begin{array}{@{}c@{}}
      \true{}\\
      \false{}\\
    \end{array} \right\}
  \Leftrightarrow
  \gamma(\, \mskol{ans}{\bar x} \,) =
  \left\{
    \begin{array}{@{}c@{}}
      \wonPos{\won}\\
      \lostPos{\lost}\\
    \end{array} \right\} 
\end{math}
\end{enumerate}

\normalfont{
\mypara{Sketch}
It is easy to see that one can associate with every non-recursive \datalogneg
program $\Q$ and input $D$ an evaluation game graph $G_{Q,D}$ together with a
solved game $\gsol_\ProgD{}{\D}$.
Since the game graph is acyclic, the solved game will not contain any
drawn positions. 
Further, by construction, $\gsol_\ProgD{}{\D}$ models query evaluation
of $\Q(\D)$.  }
\end{Theorem}

\subsection{Relationship with Provenance Polynomials -- How-Provenance for
\RAplus\label{sec:}}

Game graphs are constructed to preserve provenance information 
available in program and database. It turns out that for positive Datalog
programs $Q$ they generate semiring provenance polynomials as defined in
\cite{green2007provenance,grigoris-tj-simgodrec-2012} for atoms $A(\bar x) \in Q(D)$.

\mypara{Semiring Provenance Polynomials} Semiring provenance
\cite{green2007provenance,grigoris-tj-simgodrec-2012} attaches provenance
information to EDB and IDB facts. The provenance information are elements of a
commutative semiring $K$. A commutative semiring is an algebraic structure with
two distinct associative and commutative operations ``\KPlus'' and ``\KTimes''.
During query evaluation, result facts are annotated with elements from $K$ that
are created by combining the provenance information from input facts. For
example, in the join \rel{ R(a,b) \la S(a,b), T(a) } with \rel{S(a,b)} being
annotated with $p_1 \in K$ and \rel{T(a)} being annotated with $p_2 \in K$, the
result fact \rel{R(a,b)} will be annotated with $p_1 \KTimes p_2$. Intuitively,
``\KTimes'' is used to combine provenance information of \emph{joint} use of
input facts, whereas ``\KPlus'' is used for alternative use of input facts.

Depending on the conrete semiring used, different (provenance)
information is propagated during query evaluation. The \emph{most
  informative}\footnote{In the sense that for any other semiring $K'$,
  there exists a semiring homomorphism $\mathcal H : \NX
  \rightarrow K'$.  This has important implications in
  practice \cite{green2007provenance,grigoris-tj-simgodrec-2012}.}
semiring is the \emph{positive algebra provenance semiring}
$\NX$ \cite{green2007provenance,grigoris-tj-simgodrec-2012}
whose elements are polynomials with variables from a set $X$ and
coefficients from $\mathbb{N}$.  The operators ``\KTimes'' and
``\KPlus'' in $\NX$ are the usual addition and multiplication of
polynomials. Usually, facts from the input database $D$ are annotate
by variables from a set $X$. Formally, we use $\tj^{\NX}$ as a
function that maps a ground atom to its provenance annotation in \NX.

\mypara{Obtaining Semiring Polynomials from Game Provenance} 
Let $Q$ be a positive query, and fix an atom $A(\bar
x)\in Q(D)$.
The provenance graph $\gprov_\ProgD{}{\D}(\mskol{A}{\bar x}) = (V,M,\glabel)$ for $A(\bar x)$ can easily
be transformed into an operator tree for a provenance polynomial. 
The operator tree is represented as a DAG \TODO{remove: $G = $}$G^\OP(A(\bar x))$ in which common sub-expressions are re-used.
$G^\OP(A(\bar x)) = (V',M',\delta)$ has nodes $V'$, edges $M'$, and node labels
$\delta$. For a fixed $A(\bar x)$, the structures of $\gprov$ and $G^{\OP}$ coincide, that is
$V=V'$ and $M=M'$.
The labeling function $\delta$ maps inner nodes to either ``$+$'' or ``$\KTimes$'', 
denoting n-ary versions of the semiring operators.
Leaf nodes in game provenance graphs correspond to atoms over the EDB schema. We
here only assign elements from $K$ to leaf nodes of the form $\mskol{r_R}{\bar x}$. 
Formally, the labeling function $\delta$ is defined as follows:
\begin{equation}\label{eq:simpleL}
  \delta(v) = 
  \begin{cases}
     \tj^{\NX}(A(\bar x)) &\textnormal{if }\;\;  \Flr(v) = \emptyset
     \;\textnormal{ and }\; v = \mskol{r_A}{\bar x}\\
     \mbox{``$\KTimes$''} &\textnormal{if }\;\;  \Flr(v) \neq \emptyset
     \;\textnormal{ and }\; \glabel(v)=\lost \\ 
     \mbox{``$\KPlus$''}  &\textnormal{if }\;\;  \Flr(v) \neq \emptyset
     \;\textnormal{ and }\; \glabel(v)=\won \\ 
  \end{cases}
\end{equation}
We use $\OP$ to denote the transformation of obtaining 
$G^\OP(A(\bar x))$ from $\gprov_\ProgD{}{\D}(\mskol{A}{\bar x})$. 
The provenance semiring polynomial of fact $A(\bar x)$ 
is now explicit in $G^\OP(A(\bar x))$. An
inner node ``$+$'' (or ``$\KTimes$'') with $n$ children represents an $n$-ary
version of $+$ (or $\KTimes$) from the semiring. Since the semiring operators are associative
and commutative, their $n$-ary versions are well-defined.

\begin{Proposition}
  For positive $Q$, and $A(\bar x)\in \Q(\D)$, all leaves in
  $\gprov_\ProgD{}{\D}(A(\bar x))$ are of type $\mskol{r_B}{X,Y}$;
  thus the labeling described above is complete.

\normalfont{\mypara{Sketch}
For positive programs, positive relation nodes are reachable from other positive
relation nodes over a path of length four as shown on the left side of \figref{fig-game-overview}.
For an atom $A(\bar x)\in Q(D)$, all reachable rule nodes are lost and all reachable goal nodes are 
won.}
\end{Proposition}

 \setlength{\tabcolsep}{0.2em}
 \renewcommand{\arraystretch}{1.1}

 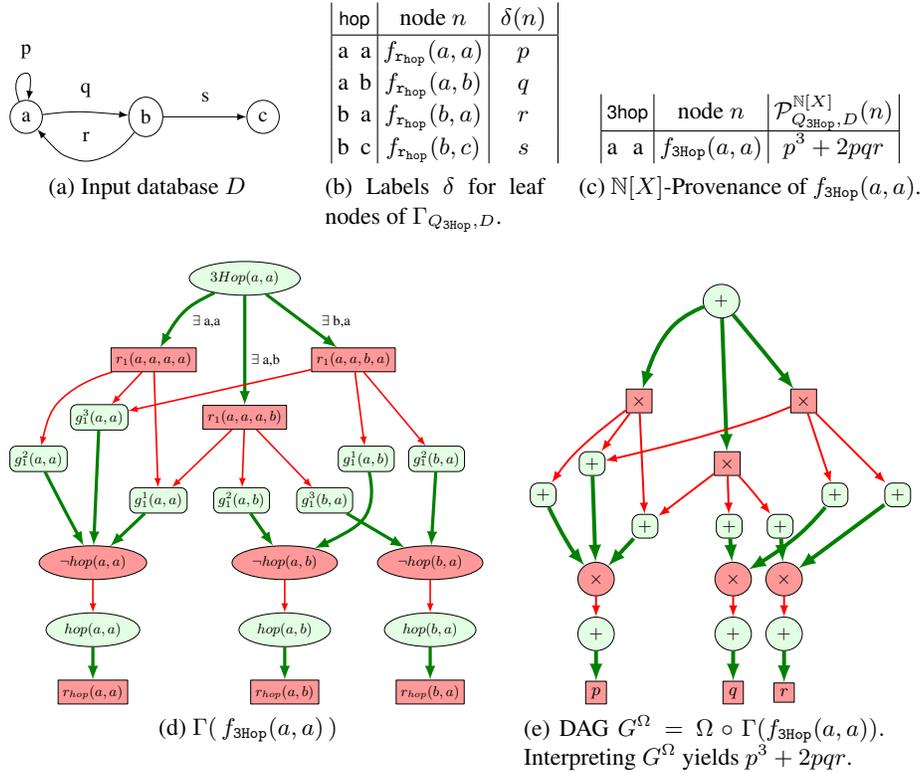
\begin{figure}[thb]
  \centering
\subfloat[Input database $D$]{
  \mbox{\resizebox{.30\columnwidth}{!}{\begin{tikzpicture}[>=latex,line join=bevel,]
\node (a) at (9bp,19.296bp) [draw,ellipse] {a};
  \node (c) at (112bp,19.296bp) [draw,ellipse] {c};
  \node (b) at (61bp,19.296bp) [draw,ellipse] {b};
  \draw [->] (b) ..controls (52.383bp,7.8131bp) and (48.473bp,4.1489bp)  .. (44bp,2.2961bp) .. controls (36.609bp,-0.76537bp) and (33.391bp,-0.76537bp)  .. (26bp,2.2961bp) .. controls (24.303bp,2.9989bp) and (22.688bp,3.9623bp)  .. (a);
  \definecolor{strokecol}{rgb}{0.0,0.0,0.0};
  \pgfsetstrokecolor{strokecol}
  \draw (35bp,10.796bp) node {r};
  \draw [->] (a) ..controls (20.308bp,20.833bp) and (23.271bp,21.136bp)  .. (26bp,21.296bp) .. controls (31.865bp,21.64bp) and (34.516bp,21.731bp)  .. (b);
  \draw (35bp,30.796bp) node {q};
  \draw [->] (a) ..controls (3.1348bp,32.309bp) and (4.5312bp,38.296bp)  .. (9bp,38.296bp) .. controls (10.396bp,38.296bp) and (11.493bp,37.711bp)  .. (a);
  \draw (9bp,47.796bp) node {p};
  \draw [->] (b) ..controls (76.843bp,19.296bp) and (85.702bp,19.296bp)  .. (c);
  \draw (87bp,27.796bp) node {s};
\end{tikzpicture}}}
  \label{hopgraph}
}
\hfill
   \subfloat[Labels $\delta$ for leaf nodes of $\gprov_\PHopD$. ] {
     \parbox[b][][t]{.22\textwidth}{  %
       \centering
       \begin{tabular}{|cc|c|c|}
       \multicolumn{2}{|c|}{\textsf{\scriptsize hop}} & node $n$ & ~$\delta(n)$~ \\\hline
       a & a & $\mskol{r_{hop}}{a,a}$ & $p$ \\
       a & b & $\mskol{r_{hop}}{a,b}$ & $q$ \\
       b & a & $\mskol{r_{hop}}{b,a}$ & $r$ \\
       b & c & $\mskol{r_{hop}}{b,c}$ & $s$ \\
       \end{tabular}
     }
     \label{3hopan}
     \label{3hopinputannotations}
   }    
   \hfill
   \subfloat[${\NX}$-Provenance of $\mskol{3Hop}{a,a}$.] {
     \parbox[b][][t]{.36\textwidth}{ %
       \centering
       \begin{tabular}{|cc|c|c|c|c|c|c|}
       \multicolumn{2}{|c|}{\textsf{\scriptsize 3hop}} & node $n$ &
       $\tj^{\NX}_\PHopD(n)$ \\\hline
        a & a & $\mskol{3Hop}{a,a}$  & $p^3+2pqr$ \\
       \end{tabular}
     }
     \label{33hopan}
   }

\subfloat[$\gprov(\,\mskol{3Hop}{a,a}\,)$]{
  \mbox{\resizebox{.52\columnwidth}{!}{\begin{tikzpicture}[>=latex,line join=bevel,]
\node (RU5AbKaKaKbKaKaZ) at (216bp,217bp) [draw,rectangle, fill=red!40, draw] {${r_1}(a,a,b,a)$};
  \node (GU5U2UhopAaKaZ) at (19bp,155bp) [draw,rectangle, fill=green!11, draw,rounded corners] {${g_1^2}(a,a)$};
  \node (!hopAbKaZ) at (264bp,90bp) [draw,ellipse, fill=red!40, draw] {${\neg{}hop}(b,a)$};
  \node (GU5U1UhopAaKaZ) at (94bp,129bp) [draw,rectangle, fill=green!11, draw,rounded corners] {${g_1^1}(a,a)$};
  \node (hopAbKaZ) at (264bp,48bp) [draw,ellipse, fill=green!11, draw] {${hop}(b,a)$};
  \node (GU5U2UhopAaKbZ) at (146bp,129bp) [draw,rectangle, fill=green!11, draw,rounded corners] {${g_1^2}(a,b)$};
  \node (GU5U3UhopAbKaZ) at (198bp,129bp) [draw,rectangle, fill=green!11, draw,rounded corners] {${g_1^3}(b,a)$};
  \node (RUhopAbKaZ) at (264bp,9bp) [draw,rectangle, fill=red!40, draw] {${r_{hop}}(b,a)$};
  \node (RU5AaKaKaKaKaKaZ) at (91bp,217bp) [draw,rectangle, fill=red!40, draw] {${r_1}(a,a,a,a)$};
  \node (RU5AaKbKaKaKbKaZ) at (148bp,181bp) [draw,rectangle, fill=red!40, draw] {${r_1}(a,a,a,b)$};
  \node (threeHopAaKaZ) at (148bp,268bp) [draw,ellipse, fill=green!11, draw] {${3Hop}(a,a)$};
  \node (GU5U3UhopAaKaZ) at (57bp,181bp) [draw,rectangle, fill=green!11, draw,rounded corners] {${g_1^3}(a,a)$};
  \node (RUhopAaKaZ) at (53bp,9bp) [draw,rectangle, fill=red!40, draw] {${r_{hop}}(a,a)$};
  \node (hopAaKbZ) at (173bp,48bp) [draw,ellipse, fill=green!11, draw] {${hop}(a,b)$};
  \node (!hopAaKaZ) at (53bp,90bp) [draw,ellipse, fill=red!40, draw] {${\neg{}hop}(a,a)$};
  \node (GU5U1UhopAaKbZ) at (224bp,155bp) [draw,rectangle, fill=green!11, draw,rounded corners] {${g_1^1}(a,b)$};
  \node (RUhopAaKbZ) at (173bp,9bp) [draw,rectangle, fill=red!40, draw] {${r_{hop}}(a,b)$};
  \node (GU5U2UhopAbKaZ) at (268bp,155bp) [draw,rectangle, fill=green!11, draw,rounded corners] {${g_1^2}(b,a)$};
  \node (!hopAaKbZ) at (173bp,90bp) [draw,ellipse, fill=red!40, draw] {${\neg{}hop}(a,b)$};
  \node (hopAaKaZ) at (53bp,48bp) [draw,ellipse, fill=green!11, draw] {${hop}(a,a)$};
  \draw [->,line width=2,green!50!black] (GU5U1UhopAaKaZ) ..controls (81.089bp,116.72bp) and (77.113bp,112.94bp)  .. (!hopAaKaZ);
  \draw [->,line width=1,red] (RU5AaKbKaKaKbKaZ) ..controls (147.42bp,165.83bp) and (147.06bp,156.5bp)  .. (GU5U2UhopAaKbZ);
  \draw [->,line width=2,green!50!black] (GU5U2UhopAbKaZ) ..controls (266.91bp,137.29bp) and (266.08bp,123.85bp)  .. (!hopAbKaZ);
  \draw [->,line width=2,green!50!black] (GU5U2UhopAaKaZ) ..controls (27.367bp,138.3bp) and (33.196bp,126.79bp)  .. (!hopAaKaZ);
  \draw [->,line width=2,green!50!black] (threeHopAaKaZ) ..controls (117.71bp,253.48bp) and (114.11bp,250.88bp)  .. (111bp,248bp) .. controls (106.82bp,244.13bp) and (103.13bp,239.21bp)  .. (RU5AaKaKaKaKaKaZ);
  \definecolor{strokecol}{rgb}{0.0,0.0,0.0};
  \pgfsetstrokecolor{strokecol}
  \draw (124bp,241bp) node {$\exists$ a,a};
  \draw [->,line width=2,green!50!black] (threeHopAaKaZ) ..controls (173.71bp,248.71bp) and (185.97bp,239.52bp)  .. (RU5AbKaKaKbKaKaZ);
  \draw (205bp,241bp) node {$\exists$ b,a};
  \draw [->,line width=2,green!50!black] (GU5U3UhopAbKaZ) ..controls (219.63bp,116.22bp) and (227.17bp,111.76bp)  .. (!hopAbKaZ);
  \draw [->,line width=1,red] (RU5AbKaKaKbKaKaZ) ..controls (218.24bp,199.6bp) and (220.04bp,185.71bp)  .. (GU5U1UhopAaKbZ);
  \draw [->,line width=1,red] (!hopAaKbZ) ..controls (173bp,75.298bp) and (173bp,72.8bp)  .. (hopAaKbZ);
  \draw [->,line width=1,red] (RU5AaKbKaKaKbKaZ) ..controls (131.41bp,165.02bp) and (120.12bp,154.15bp)  .. (GU5U1UhopAaKaZ);
  \draw [->,line width=2,green!50!black] (hopAaKbZ) ..controls (173bp,33.083bp) and (173bp,30.311bp)  .. (RUhopAaKbZ);
  \draw [->,line width=2,green!50!black] (GU5U2UhopAaKbZ) ..controls (154.27bp,117.05bp) and (156.59bp,113.71bp)  .. (!hopAaKbZ);
  \draw [->,line width=2,green!50!black] (GU5U1UhopAaKbZ) ..controls (227.77bp,138.28bp) and (228.46bp,127.81bp)  .. (224bp,120bp) .. controls (221.36bp,115.38bp) and (217.7bp,111.4bp)  .. (!hopAaKbZ);
  \draw [->,line width=1,red] (RU5AaKaKaKaKaKaZ) ..controls (91.772bp,194.35bp) and (92.707bp,166.94bp)  .. (GU5U1UhopAaKaZ);
  \draw [->,line width=1,red] (RU5AaKbKaKaKbKaZ) ..controls (163.29bp,165.1bp) and (173.6bp,154.37bp)  .. (GU5U3UhopAbKaZ);
  \draw [->,line width=1,red] (!hopAbKaZ) ..controls (264bp,75.298bp) and (264bp,72.8bp)  .. (hopAbKaZ);
  \draw [->,line width=2,green!50!black] (threeHopAaKaZ) ..controls (148bp,241.41bp) and (148bp,216.96bp)  .. (RU5AaKbKaKaKbKaZ);
  \draw (161bp,217bp) node {$\exists$ a,b};
  \draw [->,line width=1,red] (!hopAaKaZ) ..controls (53bp,75.298bp) and (53bp,72.8bp)  .. (hopAaKaZ);
  \draw [->,line width=2,green!50!black] (hopAbKaZ) ..controls (264bp,33.083bp) and (264bp,30.311bp)  .. (RUhopAbKaZ);
  \draw [->,line width=1,red] (RU5AbKaKaKbKaKaZ) ..controls (184.22bp,209.65bp) and (180.52bp,208.8bp)  .. (177bp,208bp) .. controls (145.99bp,200.93bp) and (110.44bp,192.95bp)  .. (GU5U3UhopAaKaZ);
  \draw [->,line width=1,red] (RU5AaKaKaKaKaKaZ) ..controls (50.845bp,205.12bp) and (39.156bp,198.95bp)  .. (31bp,190bp) .. controls (26.992bp,185.6bp) and (24.313bp,179.81bp)  .. (GU5U2UhopAaKaZ);
  \draw [->,line width=1,red] (RU5AaKaKaKaKaKaZ) ..controls (79.689bp,205.02bp) and (76.079bp,201.2bp)  .. (GU5U3UhopAaKaZ);
  \draw [->,line width=2,green!50!black] (GU5U3UhopAaKaZ) ..controls (55.994bp,158.12bp) and (54.818bp,131.36bp)  .. (!hopAaKaZ);
  \draw [->,line width=2,green!50!black] (hopAaKaZ) ..controls (53bp,33.083bp) and (53bp,30.311bp)  .. (RUhopAaKaZ);
  \draw [->,line width=1,red] (RU5AbKaKaKbKaKaZ) ..controls (231.05bp,199.06bp) and (243.7bp,183.97bp)  .. (GU5U2UhopAbKaZ);
\end{tikzpicture}}}
 \label{3hopprovenance} 
}
\hfill
\subfloat[DAG $G^\OP = \OP \circ \gprov(\mskol{3Hop}{a,a}).\quad\;$ Interpreting
   $G^{\OP}$ yields  $p^3+2pqr$. ]{
  \mbox{\resizebox{.42\columnwidth}{!}{\begin{tikzpicture}[>=latex,line join=bevel,]
\node (RU5AbKaKaKbKaKaZ) at (136bp,156bp) [draw,rectangle, fill=red!40, draw] {$\KTimes$};
  \node (GU5U2UhopAaKaZ) at (3bp,108bp) [draw,rectangle, fill=green!11, draw,rounded corners] {$\KPlus$};
  \node (!hopAbKaZ) at (126bp,65bp) [draw,ellipse, fill=red!40, draw] {$\KTimes$};
  \node (GU5U1UhopAaKaZ) at (55bp,92bp) [draw,rectangle, fill=green!11, draw,rounded corners] {$\KPlus$};
  \node (hopAbKaZ) at (126bp,37bp) [draw,ellipse, fill=green!11, draw] {$\KPlus$};
  \node (GU5U2UhopAaKbZ) at (98bp,92bp) [draw,rectangle, fill=green!11, draw,rounded corners] {$\KPlus$};
  \node (GU5U3UhopAbKaZ) at (124bp,92bp) [draw,rectangle, fill=green!11, draw,rounded corners] {$\KPlus$};
  \node (RUhopAbKaZ) at (126bp,7bp) [draw,rectangle, fill=red!40, draw] {$r$};
  \node (RU5AaKaKaKaKaKaZ) at (52bp,156bp) [draw,rectangle, fill=red!40, draw] {$\KTimes$};
  \node (RU5AaKbKaKaKbKaZ) at (97bp,124bp) [draw,rectangle, fill=red!40, draw] {$\KTimes$};
  \node (threeHopAaKaZ) at (94bp,207bp) [draw,ellipse, fill=green!11, draw] {$\KPlus$};
  \node (GU5U3UhopAaKaZ) at (28bp,124bp) [draw,rectangle, fill=green!11, draw,rounded corners] {$\KPlus$};
  \node (RUhopAaKaZ) at (30bp,7bp) [draw,rectangle, fill=red!40, draw] {$p$};
  \node (hopAaKbZ) at (100bp,37bp) [draw,ellipse, fill=green!11, draw] {$\KPlus$};
  \node (!hopAaKaZ) at (30bp,65bp) [draw,ellipse, fill=red!40, draw] {$\KTimes$};
  \node (GU5U1UhopAaKbZ) at (152bp,108bp) [draw,rectangle, fill=green!11, draw,rounded corners] {$\KPlus$};
  \node (RUhopAaKbZ) at (100bp,7bp) [draw,rectangle, fill=red!40, draw] {$q$};
  \node (GU5U2UhopAbKaZ) at (184bp,108bp) [draw,rectangle, fill=green!11, draw,rounded corners] {$\KPlus$};
  \node (!hopAaKbZ) at (100bp,65bp) [draw,ellipse, fill=red!40, draw] {$\KTimes$};
  \node (hopAaKaZ) at (30bp,37bp) [draw,ellipse, fill=green!11, draw] {$\KPlus$};
  \draw [->,line width=2,green!50!black] (GU5U1UhopAaKaZ) ..controls (49.328bp,85.875bp) and (45.414bp,81.647bp)  .. (!hopAaKaZ);
  \draw [->,line width=1,red] (RU5AaKbKaKaKbKaZ) ..controls (97.236bp,116.45bp) and (97.411bp,110.84bp)  .. (GU5U2UhopAaKbZ);
  \draw [->,line width=2,green!50!black] (GU5U2UhopAbKaZ) ..controls (173.16bp,99.962bp) and (152.78bp,84.854bp)  .. (!hopAbKaZ);
  \draw [->,line width=2,green!50!black] (GU5U2UhopAaKaZ) ..controls (7.3044bp,98.554bp) and (12.902bp,86.988bp)  .. (19bp,78bp) .. controls (19.058bp,77.914bp) and (19.117bp,77.829bp)  .. (!hopAaKaZ);
  \draw [->,line width=2,green!50!black] (threeHopAaKaZ) ..controls (83.216bp,203.53bp) and (74.504bp,199.84bp)  .. (69bp,194bp) .. controls (62.471bp,187.08bp) and (58.161bp,177.22bp)  .. (RU5AaKaKaKaKaKaZ);
  \definecolor{strokecol}{rgb}{0.0,0.0,0.0};
  \pgfsetstrokecolor{strokecol}
  \draw (76bp,184bp) node {$\;$};
  \draw [->,line width=2,green!50!black] (threeHopAaKaZ) ..controls (100.51bp,199.35bp) and (102.88bp,196.55bp)  .. (105bp,194bp) .. controls (112.35bp,185.16bp) and (120.56bp,175.08bp)  .. (RU5AbKaKaKbKaKaZ);
  \draw (128bp,184bp) node {$\;$};
  \draw [->,line width=2,green!50!black] (GU5U3UhopAbKaZ) ..controls (124.44bp,86.092bp) and (124.65bp,83.23bp)  .. (!hopAbKaZ);
  \draw [->,line width=1,red] (RU5AbKaKaKbKaKaZ) ..controls (139.44bp,145.68bp) and (144.04bp,131.89bp)  .. (GU5U1UhopAaKbZ);
  \draw [->,line width=1,red] (!hopAaKbZ) ..controls (100bp,57.654bp) and (100bp,54.943bp)  .. (hopAaKbZ);
  \draw [->,line width=1,red] (RU5AaKbKaKaKbKaZ) ..controls (88.284bp,117.36bp) and (75.691bp,107.76bp)  .. (GU5U1UhopAaKaZ);
  \draw [->,line width=2,green!50!black] (hopAaKbZ) ..controls (100bp,29.523bp) and (100bp,26.751bp)  .. (RUhopAaKbZ);
  \draw [->,line width=2,green!50!black] (GU5U2UhopAaKbZ) ..controls (98.438bp,86.092bp) and (98.65bp,83.23bp)  .. (!hopAaKbZ);
  \draw [->,line width=2,green!50!black] (GU5U1UhopAaKbZ) ..controls (145.82bp,100.16bp) and (139.92bp,93.153bp)  .. (134bp,88bp) .. controls (127.79bp,82.595bp) and (120.26bp,77.385bp)  .. (!hopAaKbZ);
  \draw [->,line width=1,red] (RU5AaKaKaKaKaKaZ) ..controls (52.597bp,143.27bp) and (53.655bp,120.7bp)  .. (GU5U1UhopAaKaZ);
  \draw [->,line width=1,red] (RU5AaKbKaKaKbKaZ) ..controls (102.55bp,118.12bp) and (105.5bp,114.92bp)  .. (108bp,112bp) .. controls (110.22bp,109.4bp) and (112.55bp,106.56bp)  .. (GU5U3UhopAbKaZ);
  \draw [->,line width=1,red] (!hopAbKaZ) ..controls (126bp,57.654bp) and (126bp,54.943bp)  .. (hopAbKaZ);
  \draw [->,line width=2,green!50!black] (threeHopAaKaZ) ..controls (94.661bp,188.72bp) and (95.835bp,156.24bp)  .. (RU5AaKbKaKaKbKaZ);
  \draw (103bp,156bp) node {$\;$};
  \draw [->,line width=1,red] (!hopAaKaZ) ..controls (30bp,57.654bp) and (30bp,54.943bp)  .. (hopAaKaZ);
  \draw [->,line width=2,green!50!black] (hopAbKaZ) ..controls (126bp,29.256bp) and (126bp,26.087bp)  .. (RUhopAbKaZ);
  \draw [->,line width=1,red] (RU5AbKaKaKbKaKaZ) ..controls (128.1bp,152.51bp) and (118.99bp,148.65bp)  .. (111bp,146bp) .. controls (86.642bp,137.93bp) and (57.517bp,130.77bp)  .. (GU5U3UhopAaKaZ);
  \draw [->,line width=1,red] (RU5AaKaKaKaKaKaZ) ..controls (42.784bp,149.49bp) and (28.354bp,138.84bp)  .. (18bp,128bp) .. controls (15.639bp,125.53bp) and (13.311bp,122.69bp)  .. (GU5U2UhopAaKaZ);
  \draw [->,line width=1,red] (RU5AaKaKaKaKaKaZ) ..controls (46.042bp,148.06bp) and (41.203bp,141.6bp)  .. (GU5U3UhopAaKaZ);
  \draw [->,line width=2,green!50!black] (GU5U3UhopAaKaZ) ..controls (28.382bp,112.73bp) and (29.024bp,93.778bp)  .. (!hopAaKaZ);
  \draw [->,line width=2,green!50!black] (hopAaKaZ) ..controls (30bp,29.523bp) and (30bp,26.751bp)  .. (RUhopAaKaZ);
  \draw [->,line width=1,red] (RU5AbKaKaKbKaKaZ) ..controls (146bp,146bp) and (162.5bp,129.5bp)  .. (GU5U2UhopAbKaZ);
\end{tikzpicture}}}
  \label{3hopprovenanceOmega}
}

  \caption{\small Input graph for program \PHop in (a) using edge labeling according to (b). 
  Game provenance $\gprov_\PHopD$ for the query \rel{3Hop(a,a)} on input database of (a) is shown
  in (d).
  When labeling leaf nodes according to (b), lost inner nodes by ``$\KTimes$'',
  and won inner nodes by ``$+$'' then the operator DAG $G^{\OP}$ shown in (e) is
  created. 
  This DAG represents the semiring-provenance polynomial for the query \rel{3Hop(a,a)} shown in
  (c) and \cite{grigoris-tj-simgodrec-2012}.
  }\label{3hopprov}
\end{figure}

 The following theorem relates semiring provenance polynomials to the provenance
expressions we obtain in $G^{\OP}$:

\begin{Theorem}\label{thm:poly}%
  Let $\gprov_\ProgD{}{\D}$ be the game provenance of an \RAplus query $Q$
  (in the form of a positive, non-recursive Datalog program) over database $D$.
Then  $\gprov_\ProgD{}{\D}$  represents the provenance
  polynomials \NX as follows:  for all $A(\bar x)\in Q(D)$,
 \begin{displaymath}
   \OP\circ\gprov_\ProgD{}{\D}(\,\mskol{A}{\bar x}\,) \equiv
   \tj^\NX_\ProgD{}{\D}(\,A(\bar x)\,)  .
 \end{displaymath}

\normalfont{
\mypara{Sketch} %
Our game graph construction is an extension of the graph presented in Section 4.2 of 
\cite{grigoris-tj-simgodrec-2012}. Rule nodes 
correspond to the join nodes presented in \cite{grigoris-tj-simgodrec-2012}.
Named goal nodes can be seen as labels on the edges between (goal) tuple nodes and join nodes 
and allow us to identify at which position a tuple was used in the body.
For a detailed proof, please refer to Appendix \ref{sec:Proof}.
}
\end{Theorem}

\mypara{Example \rel{\mathbf{3hop}} from \cite{grigoris-tj-simgodrec-2012}} 
$\;$ Consider the \rel{3Hop} query $\PHop$ used in Figure 7 of \cite{grigoris-tj-simgodrec-2012}:
\[
r_1: \quad \rel{3Hop}(X,Y) \,\Gets\, \rel{hop}(X,Z_1), \rel{hop}(Z_1,Z_2), \rel{hop}(Z_2,Y).
\]
The query uses an input database consisting of a single binary EDB relation
\rel{hop} representing a directed graph. It asks for pairs of nodes that are reachable
via exactly three edges(=\emph{hops}). 
An input database $D$ and $\tj^{\NX}_\PHopD$ annotations of \PHop are shown in
\figref{3hopinputannotations}.
\Figref{3hopprovenance} shows the game provenance $\gprov(\mskol{3Hop}{a,a})$ 
of fact \rel{3Hop(a,a)}. Positive won relation nodes indicate the existence of the corresponding fact 
in $\Prog{3Hop}(D)$. To obtain the provenance polynomial of fact $\mskol{3Hop}{a,a}$, 
we apply $\OP$ to $\gprov(\mskol{3Hop}{a,a})$ as shown in
\figref{3hopprovenanceOmega}: we replace inner won nodes by ``$\KTimes$'', inner
lost nodes by ``$+$'', and leaf nodes by their respective annotations from $K$
as given in \figref{3hopinputannotations} and \cite{grigoris-tj-simgodrec-2012}.
The so relabeled graph encodes the provenance equation
\[
  \OP\circ\gprov_\PHopD(\mskol{3Hop}{a,a}) = (p \KTimes p \KTimes p) \KPlus (p \KTimes q \KTimes r) \KPlus (p \KTimes q \KTimes r) 
                                       = p^3 \KPlus 2pqr
\]
which is equivalent to the annotation of provenance semiring polynomials as 
shown in \figref{33hopan} and \cite{grigoris-tj-simgodrec-2012}.

\subsection{Why-Not Game Provenance for \RAplus\label{sec:WhyNotDL}}

\begin{figure}[t]
  \centering
  \mbox{\resizebox{\columnwidth}{!}{\begin{tikzpicture}[>=latex,line join=bevel,]
\node (GU5U2UhopAcKaZ) at (239bp,93bp) [draw,rectangle, fill=red!40, draw] {$g_{1}^{2}(c,a)$};
  \node (!3HopAcKaZ) at (414bp,180bp) [draw,ellipse, fill=green!11, draw] {$\neg{}3Hop(c,a)$};
  \node (GU5U2UhopAcKcZ) at (399bp,93bp) [draw,rectangle, fill=red!40, draw] {$g_{1}^{2}(c,c)$};
  \node (GU5U1UhopAcKcZ) at (349bp,93bp) [draw,rectangle, fill=red!40, draw] {$g_{1}^{1}(c,c)$};
  \node (RU5AcKbKcKcKbKaZ) at (443bp,129bp) [draw,rectangle, fill=green!11, draw] {$r_{1}(c,a,c,b)$};
  \node (!hopAcKbZ) at (481bp,54bp) [draw,ellipse, fill=green!11, draw] {$\neg hop(c,b)$};
  \node (hopAcKaZ) at (241bp,12bp) [draw,ellipse, fill=red!40, draw] {$hop(c,a)$};
  \node (GU5U2UhopAbKbZ) at (581bp,93bp) [draw,rectangle, fill=red!40, draw] {$g_{1}^{2}(b,b)$};
  \node (!hopAaKcZ) at (155bp,54bp) [draw,ellipse, fill=green!11, draw] {$\neg hop(a,c)$};
  \node (hopAcKcZ) at (362bp,12bp) [draw,ellipse, fill=red!40, draw] {$hop(c,c)$};
  \node (GU5U1UhopAcKaZ) at (108bp,93bp) [draw,rectangle, fill=red!40, draw] {$g_{1}^{1}(c,a)$};
  \node (RU5AbKcKcKbKcKaZ) at (307bp,129bp) [draw,rectangle, fill=green!11, draw] {$r_{1}(c,a,b,c)$};
  \node (RU5AaKbKcKaKbKaZ) at (98bp,129bp) [draw,rectangle, fill=green!11, draw] {$r_{1}(c,a,a,b)$};
  \node (3HopAcKaZ) at (307bp,180bp) [draw,ellipse, fill=red!40, draw] {$3Hop(c,a)$};
  \node (hopAbKbZ) at (581bp,12bp) [draw,ellipse, fill=red!40, draw] {$hop(b,b)$};
  \node (GU5U2UhopAcKbZ) at (450bp,93bp) [draw,rectangle, fill=red!40, draw] {$g_{1}^{2}(c,b)$};
  \node (GU5U2UhopAaKcZ) at (163bp,93bp) [draw,rectangle, fill=red!40, draw] {$g_{1}^{2}(a,c)$};
  \node (RU5AaKcKcKaKcKaZ) at (168bp,129bp) [draw,rectangle, fill=green!11, draw] {$r_{1}(c,a,a,c)$};
  \node (!hopAcKcZ) at (362bp,54bp) [draw,ellipse, fill=green!11, draw] {$\neg hop(c,c)$};
  \node (hopAcKbZ) at (481bp,12bp) [draw,ellipse, fill=red!40, draw] {$hop(c,b)$};
  \node (!hopAcKaZ) at (241bp,54bp) [draw,ellipse, fill=green!11, draw] {$\neg hop(c,a)$};
  \node (GU5U1UhopAcKbZ) at (501bp,93bp) [draw,rectangle, fill=red!40, draw] {$g_{1}^{1}(c,b)$};
  \node (RU5AbKbKcKbKbKaZ) at (581bp,129bp) [draw,rectangle, fill=green!11, draw] {$r_{1}(c,a,b,b)$};
  \node (!hopAbKbZ) at (581bp,54bp) [draw,ellipse, fill=green!11, draw] {$\neg hop(b,b)$};
  \node (GU5U3UhopAcKaZ) at (291bp,93bp) [draw,rectangle, fill=red!40, draw] {$g_{1}^{3}(c,a)$};
  \node (RU5AaKaKcKaKaKaZ) at (28bp,129bp) [draw,rectangle, fill=green!11, draw] {$r_{1}(c,a,a,a)$};
  \node (RU5AbKaKcKbKaKaZ) at (512bp,129bp) [draw,rectangle, fill=green!11, draw] {$r_{1}(c,a,b,a)$};
  \node (hopAaKcZ) at (155bp,12bp) [draw,ellipse, fill=red!40, draw] {$hop(a,c)$};
  \node (RU5AcKaKcKcKaKaZ) at (238bp,129bp) [draw,rectangle, fill=green!11, draw] {$r_{1}(c,a,c,a)$};
  \node (RU5AcKcKcKcKcKaZ) at (375bp,129bp) [draw,rectangle, fill=green!11, draw] {$r_{1}(c,a,c,c)$};
  \draw [->,line width=2,green!50!black] (RU5AbKbKcKbKbKaZ) ..controls (552.03bp,115.97bp) and (539.57bp,110.36bp)  .. (GU5U1UhopAcKbZ);
  \draw [->,line width=2,green!50!black] (RU5AbKbKcKbKbKaZ) ..controls (581bp,117.88bp) and (581bp,115.17bp)  .. (GU5U2UhopAbKbZ);
  \draw [->,line width=2,green!50!black] (RU5AaKcKcKaKcKaZ) ..controls (213.46bp,117.04bp) and (236.28bp,110.75bp)  .. (GU5U3UhopAcKaZ);
  \draw [->,line width=1,red] (3HopAcKaZ) ..controls (242.87bp,173.91bp) and (205.75bp,168.73bp)  .. (174bp,160bp) .. controls (157.25bp,155.4bp) and (139.23bp,148.15bp)  .. (RU5AaKbKcKaKbKaZ);
  \definecolor{strokecol}{rgb}{0.0,0.0,0.0};
  \pgfsetstrokecolor{strokecol}
  \draw (187bp,153bp) node {$\exists$ a,b};
  \draw [->,line width=1,red] (3HopAcKaZ) ..controls (257.83bp,168.19bp) and (244.73bp,164.34bp)  .. (233bp,160bp) .. controls (219.4bp,154.97bp) and (204.78bp,148.13bp)  .. (RU5AaKcKcKaKcKaZ);
  \draw (245.5bp,153bp) node {$\exists$ a,c};
  \draw [->,line width=1,red] (GU5U1UhopAcKaZ) ..controls (130.34bp,86.014bp) and (133.77bp,84.968bp)  .. (137bp,84bp) .. controls (155.99bp,78.316bp) and (176.98bp,72.223bp)  .. (!hopAcKaZ);
  \draw [->,line width=1,red] (3HopAcKaZ) ..controls (286.2bp,165.24bp) and (282.47bp,162.55bp)  .. (279bp,160bp) .. controls (271.82bp,154.72bp) and (264.01bp,148.85bp)  .. (RU5AcKaKcKcKaKaZ);
  \draw (291.5bp,153bp) node {$\exists$ c,a};
  \draw [->,line width=2,green!50!black] (RU5AcKbKcKcKbKaZ) ..controls (407.53bp,115.42bp) and (390.7bp,108.97bp)  .. (GU5U1UhopAcKcZ);
  \draw [->,line width=1,red] (GU5U2UhopAaKcZ) ..controls (160.65bp,81.548bp) and (160.1bp,78.84bp)  .. (!hopAaKcZ);
  \draw [->,line width=1,red] (3HopAcKaZ) ..controls (328.62bp,165.35bp) and (332.46bp,162.62bp)  .. (336bp,160bp) .. controls (342.78bp,154.97bp) and (350.07bp,149.27bp)  .. (RU5AcKcKcKcKcKaZ);
  \draw (366bp,153bp) node {$\exists$ c,c};
  \draw [->,line width=2,green!50!black] (RU5AcKbKcKcKbKaZ) ..controls (445.18bp,117.8bp) and (445.73bp,114.98bp)  .. (GU5U2UhopAcKbZ);
  \draw [->,line width=2,green!50!black] (!hopAcKcZ) ..controls (362bp,39.298bp) and (362bp,36.8bp)  .. (hopAcKcZ);
  \draw [->,line width=1,red] (3HopAcKaZ) ..controls (307bp,161.95bp) and (307bp,154.62bp)  .. (RU5AbKcKcKbKcKaZ);
  \draw (320bp,153bp) node {$\exists$ b,c};
  \draw [->,line width=2,green!50!black] (RU5AcKcKcKcKcKaZ) ..controls (344.48bp,115.92bp) and (331.24bp,110.24bp)  .. (GU5U3UhopAcKaZ);
  \draw [->,line width=2,green!50!black] (RU5AbKaKcKbKaKaZ) ..controls (508.55bp,117.71bp) and (507.66bp,114.79bp)  .. (GU5U1UhopAcKbZ);
  \draw [->,line width=2,green!50!black] (!hopAcKaZ) ..controls (241bp,39.298bp) and (241bp,36.8bp)  .. (hopAcKaZ);
  \draw [->,line width=2,green!50!black] (RU5AbKcKcKbKcKaZ) ..controls (336.37bp,120.97bp) and (338.72bp,120.45bp)  .. (341bp,120bp) .. controls (396.95bp,108.88bp) and (414.74bp,115.17bp)  .. (GU5U1UhopAcKbZ);
  \draw [->,line width=1,red] (3HopAcKaZ) ..controls (351.16bp,170.1bp) and (358.27bp,168.88bp)  .. (365bp,168bp) .. controls (428.12bp,159.75bp) and (446bp,174.43bp)  .. (508bp,160bp) .. controls (524.62bp,156.13bp) and (542.25bp,148.74bp)  .. (RU5AbKbKcKbKbKaZ);
  \draw (559.5bp,153bp) node {$\exists$ b,b};
  \draw [->,line width=2,green!50!black] (!3HopAcKaZ) ..controls (366bp,180bp) and (360.26bp,180bp)  .. (3HopAcKaZ);
  \draw [->,line width=2,green!50!black] (!hopAcKbZ) ..controls (481bp,39.298bp) and (481bp,36.8bp)  .. (hopAcKbZ);
  \draw [->,line width=2,green!50!black] (RU5AaKaKcKaKaKaZ) ..controls (56.751bp,116.06bp) and (68.894bp,110.6bp)  .. (GU5U1UhopAcKaZ);
  \draw [->,line width=1,red] (3HopAcKaZ) ..controls (350.98bp,170.39bp) and (358.19bp,169.06bp)  .. (365bp,168bp) .. controls (398.12bp,162.82bp) and (407.57bp,168.5bp)  .. (440bp,160bp) .. controls (456.1bp,155.78bp) and (473.25bp,148.51bp)  .. (RU5AbKaKcKbKaKaZ);
  \draw (492bp,153bp) node {$\exists$ b,a};
  \draw [->,line width=1,red] (GU5U1UhopAcKcZ) ..controls (352.84bp,81.465bp) and (353.78bp,78.653bp)  .. (!hopAcKcZ);
  \draw [->,line width=2,green!50!black] (RU5AcKaKcKcKaKaZ) ..controls (281.45bp,114.91bp) and (303.8bp,107.66bp)  .. (GU5U1UhopAcKcZ);
  \draw [->,line width=1,red] (GU5U2UhopAcKbZ) ..controls (459.56bp,80.967bp) and (462.31bp,77.516bp)  .. (!hopAcKbZ);
  \draw [->,line width=2,green!50!black] (!hopAbKbZ) ..controls (581bp,39.298bp) and (581bp,36.8bp)  .. (hopAbKbZ);
  \draw [->,line width=1,red] (GU5U1UhopAcKbZ) ..controls (495bp,81.299bp) and (493.45bp,78.277bp)  .. (!hopAcKbZ);
  \draw [->,line width=2,green!50!black] (RU5AbKcKcKbKcKaZ) ..controls (301.94bp,117.62bp) and (300.6bp,114.59bp)  .. (GU5U3UhopAcKaZ);
  \draw [->,line width=2,green!50!black] (RU5AaKbKcKaKbKaZ) ..controls (101.14bp,117.71bp) and (101.95bp,114.79bp)  .. (GU5U1UhopAcKaZ);
  \draw [->,line width=2,green!50!black] (RU5AcKcKcKcKcKaZ) ..controls (382.76bp,117.37bp) and (385bp,114bp)  .. (GU5U2UhopAcKcZ);
  \draw [->,line width=1,red] (GU5U2UhopAbKbZ) ..controls (581bp,81.631bp) and (581bp,79.027bp)  .. (!hopAbKbZ);
  \draw [->,line width=1,red] (GU5U2UhopAcKcZ) ..controls (387.43bp,80.802bp) and (383.94bp,77.131bp)  .. (!hopAcKcZ);
  \draw [->,line width=2,green!50!black] (RU5AaKcKcKaKcKaZ) ..controls (166.44bp,117.8bp) and (166.05bp,114.98bp)  .. (GU5U2UhopAaKcZ);
  \draw [->,line width=2,green!50!black] (RU5AcKaKcKcKaKaZ) ..controls (238.31bp,117.88bp) and (238.38bp,115.17bp)  .. (GU5U2UhopAcKaZ);
  \draw [->,line width=1,red] (3HopAcKaZ) ..controls (220.66bp,173.99bp) and (136.94bp,167.22bp)  .. (107bp,160bp) .. controls (89.142bp,155.69bp) and (69.939bp,148.22bp)  .. (RU5AaKaKcKaKaKaZ);
  \draw (120bp,153bp) node {$\exists$ a,a};
  \draw [->,line width=1,red] (GU5U3UhopAcKaZ) ..controls (274.94bp,80.47bp) and (269.66bp,76.352bp)  .. (!hopAcKaZ);
  \draw [->,line width=2,green!50!black] (RU5AaKcKcKaKcKaZ) ..controls (147.35bp,116.61bp) and (139.56bp,111.94bp)  .. (GU5U1UhopAcKaZ);
  \draw [->,line width=2,green!50!black] (!hopAaKcZ) ..controls (155bp,39.298bp) and (155bp,36.8bp)  .. (hopAaKcZ);
  \draw [->,line width=1,red] (3HopAcKaZ) ..controls (356.23bp,168.33bp) and (369.31bp,164.45bp)  .. (381bp,160bp) .. controls (394bp,155.05bp) and (407.91bp,148.25bp)  .. (RU5AcKbKcKcKbKaZ);
  \draw (424bp,153bp) node {$\exists$ c,b};
  \draw [->,line width=1,red] (GU5U2UhopAcKaZ) ..controls (239.58bp,81.631bp) and (239.72bp,79.027bp)  .. (!hopAcKaZ);
  \draw [->,line width=2,green!50!black] (RU5AcKcKcKcKcKaZ) ..controls (366.54bp,117.28bp) and (364.02bp,113.8bp)  .. (GU5U1UhopAcKcZ);
\end{tikzpicture}}}
  \caption{\small Why-not provenance for \rel{3Hop(c,a)} using provenance games.}
  \label{3hopprovneg}
\end{figure}
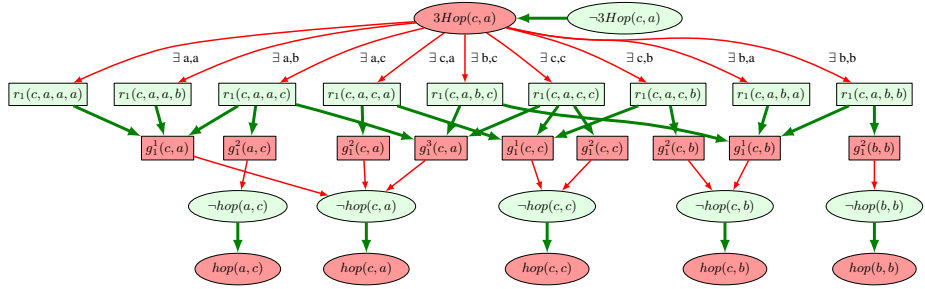

Game provenance also yields meaningful explanations for
\emph{why-not} questions. 
Consider for example the query \PHop and its input database $\D$.  The
atom $\mrel{3Hop}(c,a)$ is not in $\PHop(\D)$ and we want to get an
explanation why.  \Figref{3hopprovneg} shows the game provenance
$\gprov_\PHopD(\mskol{\neg{}3Hop}{c,a})$ of the missing fact
\rel{3Hop(c,a)}.  The lost relation node $\rel{3Hop}(c,a)$ indicates
that player I will lose the argument that tries to show that
$\rel{3Hop}(c,a) \in \Prog{3Hop}(D)$. The game provenance explains
why: Any ground instantiation of rule $r_1$ will be winning node for
player II. Consider, e.g., moving to $r_1(c,a,a,a)$ which represents
the rule instantiation for $X\mto{}c,Y\mto{}a,Z_1\mto{}a,Z_2\mto{}a$.
Player II wins the game here by questioning that the first goal
$g_1^1(c,a)$ is satisfied. And indeed, player I will move from
$g_1^1(c,a)$ to $\neg \rel{hop}(c,a)$; II to $\rel{hop}(c,a)$. Now, I
loses the game since $\rel{hop}(c,a) \not\in D$ and thus there is no
move out of $\rel{hop}(c,a)$. We also see that another rule
instantiation $X\mto{}c,Y\mto{}a,Z_1\mto{}a,Z_2\mto{}b$ fails for the
same reason: the missing $\rel{hop}(c,a)$. The instantiation
$X\mto{}c,Y\mto{}a, Z_1\mto{}b, Z_2\mto{}a$ fails because
$\rel{hop}(c,b)$ is not in the input. Other instantiations, such as
$X\mto{}c,Y\mto{}a, Z_1\mto{}c, Z_2\mto{}b$, fail because two facts
are missing from the input, here $\rel{hop}(c,b)$ and
$\rel{hop}(c,c)$.

It is no coincidence that all leaf nodes represent missing EDB facts for
why-not provenance in positive non-recursive Datalog programs:
\begin{Proposition}
Let $Q$ be a non-recursive Datalog program, $D$ a database,
$\gprov(\mskol{A}{\bar x})$ the game provenance for
facts $A(\bar x) \not\in Q(D)$. All leaves of $\gprov(\mskol{A}{\bar x})$
have  type $\mskol{R}{\bar y}$ and represent ground EDB atoms
$R(\bar y)$ that are missing from the input.
\end{Proposition}
The above proposition illustrates that for positive queries, the
ultimate reason for failure to derive outputs are missing inputs,
represented by the leaves in provenance games.

As defined, game provenance is sensitive to the active domain of query
and input database, which can lead to interesting effects.  Consider
the following query variant $\Q'_{\abc} \defeq \Pabc \cup \{ \rel{C}(y)
\la \rel E(y,z)\}$
with input $D = \{ \rel B(a,a) \}$. 
Here, game provenance shows
that $\rel A(a)$ depends on the presence of $\rel B(a,a)$ as well as on the
absence of $\rel E(a,a)$. The game provenance graph does not mention that the
absence, e.g.,  of $\rel E(a,b)$ is important as well---simply because $b$ is not in the active
domain. 

\subsection{Game Provenance for First-Order Queries}\label{datalogneg-provenance}

In this section, we demonstrate examples for provenance games in the presence of
negation within the query. When constructing game graphs for \datalogneg queries with negated goals,
we obtain graphs in which there exists a path of length three between positive
relation nodes. This switches roles between player I and II. In other words, to
explain why a \emph{negated} subgoal is satisfied, an argument like in the
why-not case is used. In general, this leads to provenance graphs that contain
leaf nodes of both kinds: $\mskol{C}{\bar x}$ representing missing facts 
$\rel{R}(\bar x) \not\in D$ and $\mskol{r_R}{\bar x}$ representing input facts 
$\rel{R}(\bar x) \in D$. 

In the following, we provide examples based on the \Pabc query (cf.
\figref{pabcgraph}) with input database 
$\D = \left\lbrace B(a,b), B(b,a),
C(a)\right\rbrace$.

\begin{figure}[t]
  \centering
  \subfloat[Game provenance graph 
  $\gprov(\mskol{A}{a})$ 
  for $A(a) \in
  \Pabc(D)$.]{
    \mbox{\resizebox{.9\columnwidth}{!}{\begin{tikzpicture}[>=latex,line join=bevel,]
\node (r1b2_cb) at (194bp,13bp) [draw=black,rectangle, fill=green!11, draw] {$g_1^2$:$\neg$C(b)};
  \node (c_b) at (266bp,12bp) [draw,ellipse, fill=red!40, draw] {C(b)};
  \node (n_b_ab) at (266bp,40bp) [draw=black,ellipse, fill=red!40, draw] {$\neg$B(a,b)};
  \node (a_a) at (19bp,26bp) [draw,ellipse, fill=green!11, draw] {A(a)};
  \node (b_ab) at (339bp,40bp) [draw,ellipse, fill=green!11, draw] {B(a,b)};
  \node (r1r_ab) at (111bp,26bp) [draw,rectangle, fill=red!40, draw] {$r_1$:[B(a,b),$\neg$C(b)]};
  \node (r1b1_ab) at (194bp,36bp) [draw=black,rectangle, fill=green!11, draw] {$g_1^1$:B(a,b)};
  \node (rBr_ab) at (403bp,40bp) [draw,rectangle, fill=red!40, draw] {$r_B$:(a,b)};
  \draw [->,line width=1,red] (b_ab) ..controls (366.62bp,40bp) and (369bp,40bp)  .. (rBr_ab);
  \draw [->,line width=2,green!50!black] (a_a) ..controls (44.503bp,26bp) and (52.032bp,26bp)  .. (r1r_ab);
  \definecolor{strokecol}{rgb}{0.0,0.0,0.0};
  \pgfsetstrokecolor{strokecol}
  \draw (54bp,31.5bp) node {$\exists$b};
  \draw [->,line width=1,red] (n_b_ab) ..controls (298.32bp,40bp) and (300.99bp,40bp)  .. (b_ab);
  \draw [->,line width=1,red] (r1r_ab) ..controls (154.73bp,31.269bp) and (157.45bp,31.596bp)  .. (r1b1_ab);
  \draw [->,line width=2,green!50!black] (r1b2_cb) ..controls (223.57bp,12.589bp) and (230.08bp,12.499bp)  .. (c_b);
  \draw [->,line width=1,red] (r1r_ab) ..controls (154.73bp,19.151bp) and (157.45bp,18.725bp)  .. (r1b2_cb);
  \draw [->,line width=2,green!50!black] (r1b1_ab) ..controls (220.38bp,37.466bp) and (223.3bp,37.628bp)  .. (n_b_ab);
\end{tikzpicture}}}
    \label{provgamegraphpabcAa}
  }
    
  \subfloat[Game provenance graph 
    $\gprov(\mskol{\neg{}A}{b})$ for $A(b) \not\in \Pabc(D)$    
    ]{  
    \mbox{\resizebox{.9\columnwidth}{!}{\begin{tikzpicture}[>=latex,line join=bevel,]
\node (c_a) at (350bp,40bp) [draw,ellipse, fill=green!11, draw] {C(a)};
  \node (r1b2_ca) at (278bp,38bp) [draw=black,rectangle, fill=red!40, draw] {$g_1^2$:$\neg$C(a)};
  \node (r1b1_bb) at (278bp,13bp) [draw=black,rectangle, fill=red!40, draw] {$g_1^1$:B(b,b)};
  \node (n_a_b) at (25bp,24bp) [draw=black,ellipse, fill=green!11, draw] {$\neg$A(b)};
  \node (n_b_bb) at (350bp,12bp) [draw=black,ellipse, fill=green!11, draw] {$\neg$B(b,b)};
  \node (rCr_a) at (423bp,40bp) [draw,rectangle, fill=red!40, draw] {$r_C$:(a)};
  \node (r1r_bb) at (194bp,13bp) [draw,rectangle, fill=green!11, draw] {$r_1$:[B(b,b),$\neg$C(b)]};
  \node (r1r_ba) at (194bp,34bp) [draw,rectangle, fill=green!11, draw] {$r_1$:[B(b,a),$\neg$C(a)]};
  \node (a_b) at (100bp,24bp) [draw,ellipse, fill=red!40, draw] {A(b)};
  \node (b_bb) at (423bp,12bp) [draw,ellipse, fill=red!40, draw] {B(b,b)};
  \draw [->,line width=1,red] (r1b1_bb) ..controls (304.75bp,12.628bp) and (307.2bp,12.594bp)  .. (n_b_bb);
  \draw [->,line width=2,green!50!black] (r1r_bb) ..controls (238.26bp,13bp) and (241.01bp,13bp)  .. (r1b1_bb);
  \draw [->,line width=2,green!50!black] (n_a_b) ..controls (56.062bp,24bp) and (63.158bp,24bp)  .. (a_b);
  \definecolor{strokecol}{rgb}{0.0,0.0,0.0};
  \pgfsetstrokecolor{strokecol}
  \draw (65bp,33.5bp) node { };
  \draw [->,line width=1,red] (c_a) ..controls (377.23bp,40bp) and (386.74bp,40bp)  .. (rCr_a);
  \draw [->,line width=1,red] (a_b) ..controls (122.47bp,16.139bp) and (125.28bp,15.468bp)  .. (128bp,15bp) .. controls (132.51bp,14.223bp) and (137.21bp,13.64bp)  .. (r1r_bb);
  \draw (136bp,20.5bp) node {$\exists$b};
  \draw [->,line width=2,green!50!black] (n_b_bb) ..controls (382.57bp,12bp) and (385.06bp,12bp)  .. (b_bb);
  \draw [->,line width=2,green!50!black] (r1r_ba) ..controls (238.32bp,36.111bp) and (241.64bp,36.269bp)  .. (r1b2_ca);
  \draw [->,line width=1,red] (r1b2_ca) ..controls (307.29bp,38.814bp) and (313.96bp,38.999bp)  .. (c_a);
  \draw [->,line width=1,red] (a_b) ..controls (126.75bp,26.846bp) and (134.73bp,27.694bp)  .. (r1r_ba);
  \draw (136bp,33.5bp) node {$\exists$a};
\end{tikzpicture}}}
    \label{provgamegraphpabcAb}
  }

  \caption{\small \label{provgamegraphspabc} Provenance graphs for $\Pabc$ with
  database $\D= \{ B(a,b), B(b,a), C(a)\}$. Both why and why-not graphs might
  contain leaf nodes representing existent and missing input facts.}
\end{figure}
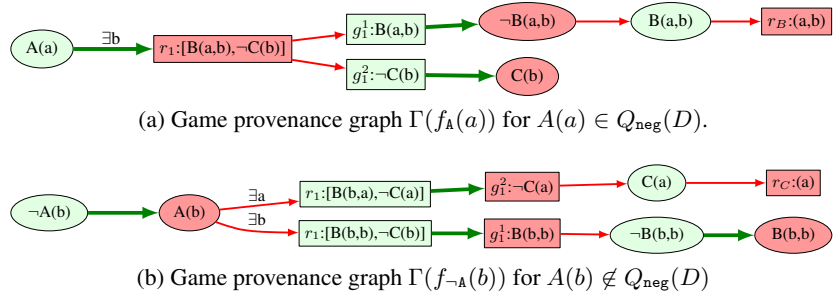

\mypara{Why Provenance} \Figref{provgamegraphpabcAa} shows the provenance graph
for the output fact $A(a)$. One can see that $A(a)$ could be derived via rule
$r_1$ with the bindings $X\mto{}a,Y\mto{}b$. The positive goal succeeds due to the
existence of the EDB fact $B(a,b)$. The negative goal $g_1^2$ succeeds due to
the missing fact $C(b)$ from the input $D$.

\mypara{Why-Not Provenance} \Figref{provgamegraphpabcAb} shows the provenance
graph for $A(b)$ which is not part of $\Pabc(D)$. We can see that a player
starting in $\neg A(b)$ will win the argument since $A(b)$ cannot be shown. 
Both attempts to derive $A(b)$ fail. With $X\mto{}b,Y\mto{}a$ the second goal $\neg C(a)$
is not satisfied since $C(a) \in D$. With $X\mto{}b,Y\mto{}b$ the first goal $B(b,b)$
fails since $B(b,b) \not\in D$.

\subsection{Evaluation Game Graph Variants}

In the graph construction for provenance games, the definition of the Skolem functions
is critical to capture provenance equivalent to \NX povenance polynomials. 
Recall that the Skolem function for rule node identifiers, \eg
$\mskol{r_1}{X,Y}$, depend on the rule (here $\mathtt{r}_1$) as well as the constants assigned to
body variables. Skolem functions of goal node identifiers, \eg
$\mskol{g_1^2}{X,Y}$, depend on the rule they belong to (here 1), 
the exact position in the rule body at which that goal oocurs (here 2), 
and values of the bound variables.

By changing the definition of one or more Skolem functions, more compact but also less 
informative provenance can be encoded. We here only describe a simple variant
that will create \TrioProv \cite{benjelloun2006uldbs} style provenance instead of \NX provenance
polynomials for \RAplus queries.
When changing the Skolem function of goal node identifiers by removing the
positional argument for the goal, goals that appear at different positions in
the body of a rule collapse into a single node. This construction yields a modified operator graph. 
In particular, using the same fact multiple times jointly in a rule will be recorded only as a
single use---as it is the case in \TrioProv provenance polynomials. 

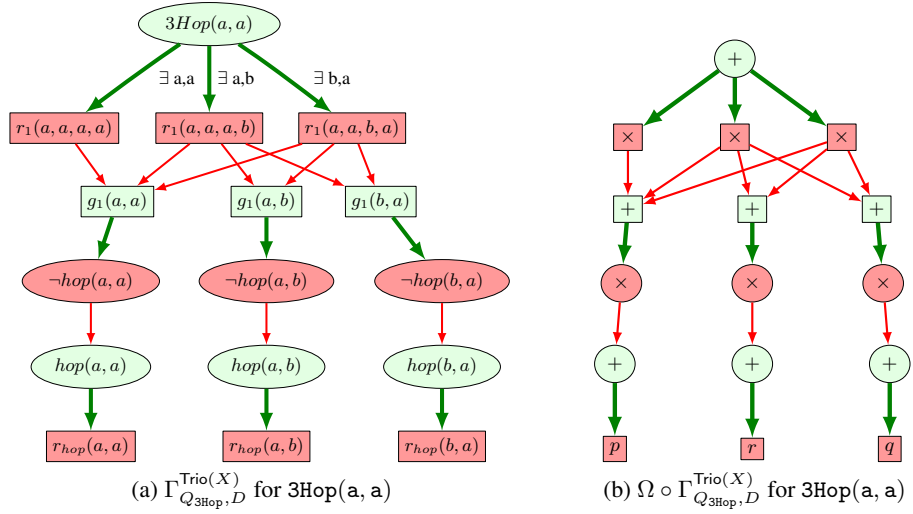
\begin{figure}[t]
  \subfloat[$\gprov_\PHopD^\TrioProv$ for \rel{3Hop(a,a)}]{
    \mbox{\resizebox{.55\columnwidth}{!}{\begin{tikzpicture}[>=latex,line join=bevel,]
\node (RU5AbKaKaKbKaKaZ) at (170bp,165bp) [draw,rectangle, fill=red!40, draw] {${r_1}(a,a,b,a)$};
  \node (GU5UhopAbKaZ) at (184bp,129bp) [draw,rectangle, fill=green!11, draw] {${g_1}(b,a)$};
  \node (GU5UhopAaKbZ) at (128bp,129bp) [draw,rectangle, fill=green!11, draw] {${g_1}(a,b)$};
  \node (RUhopAbKaZ) at (214bp,9bp) [draw,rectangle, fill=red!40, draw] {${r_{hop}}(b,a)$};
  \node (!hopAbKaZ) at (214bp,90bp) [draw,ellipse, fill=red!40, draw] {${\neg{}hop}(b,a)$};
  \node (RU5AaKaKaKaKaKaZ) at (29bp,165bp) [draw,rectangle, fill=red!40, draw] {${r_1}(a,a,a,a)$};
  \node (RU5AaKbKaKaKbKaZ) at (100bp,165bp) [draw,rectangle, fill=red!40, draw] {${r_1}(a,a,a,b)$};
  \node (threeHopAaKaZ) at (100bp,216bp) [draw,ellipse, fill=green!11, draw] {${3Hop}(a,a)$};
  \node (RUhopAaKaZ) at (42bp,9bp) [draw,rectangle, fill=red!40, draw] {${r_{hop}}(a,a)$};
  \node (hopAaKbZ) at (128bp,48bp) [draw,ellipse, fill=green!11, draw] {${hop}(a,b)$};
  \node (!hopAaKaZ) at (42bp,90bp) [draw,ellipse, fill=red!40, draw] {${\neg{}hop}(a,a)$};
  \node (GU5UhopAaKaZ) at (55bp,129bp) [draw,rectangle, fill=green!11, draw] {${g_1}(a,a)$};
  \node (RUhopAaKbZ) at (128bp,9bp) [draw,rectangle, fill=red!40, draw] {${r_{hop}}(a,b)$};
  \node (hopAbKaZ) at (214bp,48bp) [draw,ellipse, fill=green!11, draw] {${hop}(b,a)$};
  \node (!hopAaKbZ) at (128bp,90bp) [draw,ellipse, fill=red!40, draw] {${\neg{}hop}(a,b)$};
  \node (hopAaKaZ) at (42bp,48bp) [draw,ellipse, fill=green!11, draw] {${hop}(a,a)$};
  \draw [->,line width=2,green!50!black] (threeHopAaKaZ) ..controls (121.57bp,201.29bp) and (125.43bp,198.58bp)  .. (129bp,196bp) .. controls (136.22bp,190.78bp) and (144.05bp,184.92bp)  .. (RU5AbKaKaKbKaKaZ);
  \definecolor{strokecol}{rgb}{0.0,0.0,0.0};
  \pgfsetstrokecolor{strokecol}
  \draw (160bp,189bp) node {$\exists$ b,a};
  \draw [->,line width=2,green!50!black] (threeHopAaKaZ) ..controls (78.447bp,201.26bp) and (74.584bp,198.56bp)  .. (71bp,196bp) .. controls (63.634bp,190.74bp) and (55.626bp,184.87bp)  .. (RU5AaKaKaKaKaKaZ);
  \draw (84bp,189bp) node {$\exists$ a,a};
  \draw [->,line width=1,red] (RU5AaKbKaKaKbKaZ) ..controls (130.52bp,151.92bp) and (143.76bp,146.24bp)  .. (GU5UhopAbKaZ);
  \draw [->,line width=2,green!50!black] (GU5UhopAbKaZ) ..controls (192.99bp,117.32bp) and (195.71bp,113.77bp)  .. (!hopAbKaZ);
  \draw [->,line width=1,red] (RU5AaKbKaKaKbKaZ) ..controls (109.25bp,153.1bp) and (112.15bp,149.38bp)  .. (GU5UhopAaKbZ);
  \draw [->,line width=1,red] (RU5AbKaKaKbKaKaZ) ..controls (125.01bp,150.92bp) and (101.9bp,143.68bp)  .. (GU5UhopAaKaZ);
  \draw [->,line width=1,red] (!hopAbKaZ) ..controls (214bp,75.298bp) and (214bp,72.8bp)  .. (hopAbKaZ);
  \draw [->,line width=1,red] (RU5AaKaKaKaKaKaZ) ..controls (37.529bp,153.19bp) and (40.129bp,149.59bp)  .. (GU5UhopAaKaZ);
  \draw [->,line width=1,red] (!hopAaKaZ) ..controls (42bp,75.298bp) and (42bp,72.8bp)  .. (hopAaKaZ);
  \draw [->,line width=1,red] (!hopAaKbZ) ..controls (128bp,75.298bp) and (128bp,72.8bp)  .. (hopAaKbZ);
  \draw [->,line width=2,green!50!black] (hopAbKaZ) ..controls (214bp,33.083bp) and (214bp,30.311bp)  .. (RUhopAbKaZ);
  \draw [->,line width=2,green!50!black] (threeHopAaKaZ) ..controls (100bp,197.95bp) and (100bp,190.62bp)  .. (RU5AaKbKaKaKbKaZ);
  \draw (113bp,189bp) node {$\exists$ a,b};
  \draw [->,line width=1,red] (RU5AbKaKaKbKaKaZ) ..controls (174.46bp,153.54bp) and (175.68bp,150.41bp)  .. (GU5UhopAbKaZ);
  \draw [->,line width=2,green!50!black] (hopAaKbZ) ..controls (128bp,33.083bp) and (128bp,30.311bp)  .. (RUhopAaKbZ);
  \draw [->,line width=1,red] (RU5AbKaKaKbKaKaZ) ..controls (155.6bp,152.66bp) and (150.56bp,148.34bp)  .. (GU5UhopAaKbZ);
  \draw [->,line width=1,red] (RU5AaKbKaKaKbKaZ) ..controls (84.464bp,152.57bp) and (78.913bp,148.13bp)  .. (GU5UhopAaKaZ);
  \draw [->,line width=2,green!50!black] (hopAaKaZ) ..controls (42bp,33.083bp) and (42bp,30.311bp)  .. (RUhopAaKaZ);
  \draw [->,line width=2,green!50!black] (GU5UhopAaKbZ) ..controls (128bp,117.9bp) and (128bp,115.13bp)  .. (!hopAaKbZ);
  \draw [->,line width=2,green!50!black] (GU5UhopAaKaZ) ..controls (51.245bp,117.74bp) and (50.249bp,114.75bp)  .. (!hopAaKaZ);
\end{tikzpicture}}}
    \label{3hopprovt}
  }
  \hfill
  \subfloat[$\OP \circ \gprov_\PHopD^\TrioProv$ for \rel{3Hop(a,a)}]{
    \mbox{\resizebox{.35\columnwidth}{!}{\begin{tikzpicture}[>=latex,line join=bevel,]
\node (RU5AbKaKaKbKaKaZ) at (65bp,150bp) [draw,rectangle, fill=red!40, draw] {$\KTimes$};
  \node (GU5UhopAbKaZ) at (73bp,117bp) [draw,rectangle, fill=green!11, draw] {$+$};
  \node (GU5UhopAaKbZ) at (130bp,117bp) [draw,rectangle, fill=green!11, draw] {$+$};
  \node (RUhopAbKaZ) at (73bp,7bp) [draw,rectangle, fill=red!40, draw] {$r$};
  \node (!hopAbKaZ) at (73bp,83bp) [draw,ellipse, fill=red!40, draw] {$\KTimes$};
  \node (RU5AaKaKaKaKaKaZ) at (16bp,150bp) [draw,rectangle, fill=red!40, draw] {$\KTimes$};
  \node (RU5AaKbKaKaKbKaZ) at (114bp,150bp) [draw,rectangle, fill=red!40, draw] {$\KTimes$};
  \node (threeHopAaKaZ) at (65bp,186bp) [draw,ellipse, fill=green!11, draw] {$+$};
  \node (RUhopAaKaZ) at (10bp,7bp) [draw,rectangle, fill=red!40, draw] {$p$};
  \node (hopAaKbZ) at (136bp,46bp) [draw,ellipse, fill=green!11, draw] {$+$};
  \node (!hopAaKaZ) at (13bp,83bp) [draw,ellipse, fill=red!40, draw] {$\KTimes$};
  \node (GU5UhopAaKaZ) at (16bp,117bp) [draw,rectangle, fill=green!11, draw] {$+$};
  \node (RUhopAaKbZ) at (136bp,7bp) [draw,rectangle, fill=red!40, draw] {$q$};
  \node (hopAbKaZ) at (73bp,46bp) [draw,ellipse, fill=green!11, draw] {$+$};
  \node (!hopAaKbZ) at (133bp,83bp) [draw,ellipse, fill=red!40, draw] {$\KTimes$};
  \node (hopAaKaZ) at (10bp,46bp) [draw,ellipse, fill=green!11, draw] {$+$};
  \draw [->,line width=2,green!50!black] (threeHopAaKaZ) ..controls (65bp,172.3bp) and (65bp,167.8bp)  .. (RU5AbKaKaKbKaKaZ);
  \draw [->,line width=2,green!50!black] (threeHopAaKaZ) ..controls (46.619bp,172.5bp) and (35.595bp,164.4bp)  .. (RU5AaKaKaKaKaKaZ);
  \draw [->,line width=1,red] (RU5AaKbKaKaKbKaZ) ..controls (105.91bp,143.49bp) and (96.336bp,135.78bp)  .. (GU5UhopAbKaZ);
  \draw [->,line width=2,green!50!black] (GU5UhopAbKaZ) ..controls (73bp,106.49bp) and (73bp,102.45bp)  .. (!hopAbKaZ);
  \draw [->,line width=1,red] (RU5AaKbKaKaKbKaZ) ..controls (117.28bp,143.23bp) and (119.73bp,138.18bp)  .. (GU5UhopAaKbZ);
  \draw [->,line width=1,red] (RU5AbKaKaKbKaKaZ) ..controls (55.791bp,143.8bp) and (42.545bp,134.88bp)  .. (GU5UhopAaKaZ);
  \draw [->,line width=1,red] (!hopAbKaZ) ..controls (73bp,74.639bp) and (73bp,70.268bp)  .. (hopAbKaZ);
  \draw [->,line width=1,red] (RU5AaKaKaKaKaKaZ) ..controls (16bp,143.48bp) and (16bp,138.94bp)  .. (GU5UhopAaKaZ);
  \draw [->,line width=1,red] (!hopAaKaZ) ..controls (12.322bp,74.639bp) and (11.968bp,70.268bp)  .. (hopAaKaZ);
  \draw [->,line width=1,red] (!hopAaKbZ) ..controls (133.68bp,74.639bp) and (134.03bp,70.268bp)  .. (hopAaKbZ);
  \draw [->,line width=2,green!50!black] (hopAbKaZ) ..controls (73bp,32.111bp) and (73bp,27.321bp)  .. (RUhopAbKaZ);
  \draw [->,line width=2,green!50!black] (threeHopAaKaZ) ..controls (83.381bp,172.5bp) and (94.405bp,164.4bp)  .. (RU5AaKbKaKaKbKaZ);
  \draw [->,line width=1,red] (RU5AbKaKaKbKaKaZ) ..controls (66.602bp,143.39bp) and (67.741bp,138.69bp)  .. (GU5UhopAbKaZ);
  \draw [->,line width=2,green!50!black] (hopAaKbZ) ..controls (136bp,32.418bp) and (136bp,28.044bp)  .. (RUhopAaKbZ);
  \draw [->,line width=1,red] (RU5AbKaKaKbKaKaZ) ..controls (76.206bp,144.31bp) and (98.15bp,133.17bp)  .. (GU5UhopAaKbZ);
  \draw [->,line width=1,red] (RU5AaKbKaKaKbKaZ) ..controls (98.755bp,144.87bp) and (56.604bp,130.67bp)  .. (GU5UhopAaKaZ);
  \draw [->,line width=2,green!50!black] (hopAaKaZ) ..controls (10bp,32.418bp) and (10bp,28.044bp)  .. (RUhopAaKaZ);
  \draw [->,line width=2,green!50!black] (GU5UhopAaKbZ) ..controls (130.93bp,106.49bp) and (131.28bp,102.45bp)  .. (!hopAaKbZ);
  \draw [->,line width=2,green!50!black] (GU5UhopAaKaZ) ..controls (15.073bp,106.49bp) and (14.716bp,102.45bp)  .. (!hopAaKaZ);
\end{tikzpicture}}}
    \label{3hopprovtop}
  }
  \caption{\label{3hopprovmod}\small Creating \TrioProv style provenance game variants for $\Prog{3Hop}$
           by dropping positional identifiers in the Skolem function for goal
           nodes. The operator tree on the right reads $p+ 2 pqr$.}
\end{figure}

The game graph $\gprov_\PHopD^\TrioProv\left(\mskol{3Hop}{a,a}\right)$ and the corresponding operator graph 
are shown in \figref{3hopprovmod}. Reading out the polynomial results in the
Trio-provenance-polynomial $p + 2pqr$ for the input fact annotations given in
\figref{3hopinputannotations}.

\section{Conclusions}\label{sec-conclusions}

In this paper, we first tried to answer the question: What is the
provenance of answers to the game query $Q_G$? This non-stratified
query consists of a single rule:
\begin{equation}
\pos{win}(\bar X) \la \pos{move}(\bar X, \bar Y), \neg \pos{win}(\bar Y) \tag{$Q_G$}   
\end{equation}
To answer the question, we have proposed a natural and intuitive
notion of \emph{game provenance}, which is derived from basic
game-theoretic properties of solved games: The value and provenance of
a position $x$ depends only on a certain subgraph $\gprov(x)$ of
``good'' moves, reachable from $x$, but is independent of ``bad''
moves. $\gprov(x)$ has an elegant regular structure, i.e., alternating
\emph{winning} and \emph{delaying} moves for positions that are won or
lost, and \emph{drawing} moves for positions that are neither. 

Since $Q_G$ is a normal form for fixpoint logic
\cite{kubierschky1995remisfreie,flum-ICDT-97,flum-TCS-00}, all
fixpoint queries (and thus all first-order queries FO) can be
expressed as win-move games. Inspired by the reduction of query
evaluation to games in \cite{flum-ICDT-97}, we then sought to answer
the question: Can we use game provenance and apply it to query
evaluation games, thus hopefully obtaining a useful provenance model
for FO queries? It turns out, we can: First-order queries, expressed
as non-recursive \datalogneg programs, can be evaluated using a simple
and elegant game that resembles the well-known SLD resolution. For
positive queries our game provenance coincides with semiring
provenance. Moreover, game provenance (unlike semiring provenance)
naturally extends to full first-order queries with negation. In
particular, a simple form of \emph{why-not} provenance results from
our use of a game-theoretic semantics for querying.\footnote{See
  \cite{hintikka1996principles}, \cite{sep-logic-games}, and
  \cite{graedel11:_back_forth_between_logic_games} for other uses of
  game-theory for query evaluation and model checking.}

\mypara{Acknowledgments} Work supported in part by NSF awards
IIS-1118088, DBI-1147273, and a gift from LogicBlox, Inc.

\bibliographystyle{splncs03}
\appendix

\section{Proof of Theorem \ref{thm:poly}}\label{sec:Proof}

\begin{Proof}
The evaluation of the transformed game graph $\OP\circ\gprov_\ProgD{}{\D}(\,\mskol{R}{\bar x}\,)$
is structurally equivalent to the evaluation of provenance semiring polynomials of the annotated 
$\Q(\D)$:

\mysubpara{EDB Facts} Using provenance semirings, a fact $\rel R(\bar x)$ has 
the annotation $\tj^\NX_\ProgD{}{\D}(\,R(\bar x)\,)$. 
The evaluation of provenance polynomials 
using provenance games starts at the positive relation node $\mskol{R}{\bar x}$.
Since $R(\bar x)\in \Q(\D)$ and by definition of the game graph this relation node has one reachable node 
$\Flr(\mskol{R}{\bar x}) = \{ \mskol{r_R}{\bar x} \}$:
$\OP\circ\gprov_\ProgD{}{\D}(\mskol{R}{\bar x}) = \OP\circ\gprov_\ProgD{}{\D}(\mskol{r_R}{\bar x})$.
The node $\mskol{r_R}{\bar x}$ is a leaf node, so the evaluation \OP\ returns its label 
$L(\mskol{r_R}{\bar x}) = \tj^\NX_\ProgD{}{\D}(R(\bar x))$ and we have:
\[
  \OP\circ\gprov_\ProgD{}{\D}(\mskol{R}{\bar x}) = L(\mskol{r_R}{\bar x}) = \tj^\NX_\ProgD{}{\D}(R(\bar x)). 
\]

\mysubpara{Union} Let 
$\Q(\D) := \{ r_1\colon U(\bar x) \gets R_1(\bar x).\,\, r_2\colon U(\bar x) \gets R_2(\bar x).\} $ 
When evaluating $\Q(\D)$, the provenance semiring polynomial for fact $U(\bar x)\in \Q(\D)$ is: 
$\tj^\NX_\ProgD{}{\D}(U(\bar x)) = \tj^\NX_\ProgD{}{\D}(R_1(\bar x)) \KPlus \tj^\NX_\ProgD{}{\D}(R_2(\bar x))$.
The evaluation of provenance polynomials for $U(\bar x)\in \Q(\D)$
using provenance games starts at the positive relation node $\mskol{U}{\bar x}$.
By definition of the game graph for $\Q(\D)$, 
$\Flr(\mskol{U}{\bar x}) = \{ \mskol{r_1}{\bar x}, \mskol{r_2}{\bar x} \}$ 
and since $\glabel(\mskol{U}{\bar x})=\won$
we combine both terms with $L(\mskol{U}{\bar x}) = $``\KPlus'':
\[
  \begin{array}{rcl}
     \OP\circ\gprov_\ProgD{}{\D}(\mskol{U}{\bar x}) & = & \OP\circ\gprov_\ProgD{}{\D}(\mskol{r_1}{\bar x}) \KPlus 
     \OP\circ\gprov_\ProgD{}{\D}(\mskol{r_2}{\bar x}) \\
  \end{array}
\]
Each rule node in $\gprov_\ProgD{}{\D}$ has exactly one outgoing edge to a goal node.
Since the program is positive, each goal node has exactly one following negated relation node.
Those negated relation nodes in turn have exactly one corresponding positive relation node.
As shown above for EDB facts, for positive programs and a head node $U(\bar x)\in \Q(\D)$, 
positive relation nodes lead to the corresponding provenance annotations:
\[
  \begin{array}{rcl}
     \OP\circ\gprov_\ProgD{}{\D}(\mskol{U}{\bar x}) 
                   & = & \OP\circ\gprov_\ProgD{}{\D}(\mskol{g^1_1}{\bar x}) \KPlus 
                         \OP\circ\gprov_\ProgD{}{\D}(\mskol{g^1_2}{\bar x}) \\
                   & = & \OP\circ\gprov_\ProgD{}{\D}(\mskol{\neg R_1}{\bar x}) \KPlus 
                         \OP\circ\gprov_\ProgD{}{\D}(\mskol{\neg R_2}{\bar x}) \\
                   & = & \OP\circ\gprov_\ProgD{}{\D}(\mskol{R_1}{\bar x}) \KPlus 
                         \OP\circ\gprov_\ProgD{}{\D}(\mskol{R_2}{\bar x}) \\
                   & = & \tj^\NX_\ProgD{}{\D}(R_1(\bar x)) \KPlus \tj^\NX_\ProgD{}{\D}(R_2(\bar x)) \\
  \end{array}
\]

\mysubpara{Join} Let 
$\Q(\D) := \{ r_1\colon J(\bar x) \gets R_1(\bar x), R_2(\bar x). \} $ 
When evaluating $\Q(\D)$ for a $J(\bar x)\in \Q(\D)$ using provenance semiring annotations we get: 
$\tj^\NX_\ProgD{}{\D}(J(\bar x)) = \tj^\NX_\ProgD{}{\D}(R_1(\bar x)) \KTimes \tj^\NX_\ProgD{}{\D}(R_2(\bar x))$.
The evaluation of provenance polynomials for $J(\bar x)\in \Q(\D)$
using provenance games starts at the positive relation node $\mskol{J}{\bar x}$.
By definition of the game graph for $\Q(\D)$, $\mskol{J}{\bar x}$ connects to exactly 
one rule node: $\Flr(\mskol{J}{\bar x})=\{\mskol{r_1}{\bar x}\}$. 
This rule node in turn leads to two goal nodes
$\Flr(\mskol{r_1}{\bar x}) = \{ \mskol{g^1_1}{\bar x}, \mskol{g^2_1}{\bar x} \}$, which
we combine with $L(\mskol{r_1}{\bar x}) = $``\KTimes'', since $\glabel(\mskol{r_1}{\bar x})=\lost$:
\[
  \begin{array}{rcl}
     \OP\circ\gprov_\ProgD{}{\D}(\mskol{J}{\bar x}) 
                   & = & \OP\circ\gprov_\ProgD{}{\D}(\mskol{r_1}{\bar x}) \\
                   & = & \OP\circ\gprov_\ProgD{}{\D}(\mskol{g^1_1}{\bar x}) \KTimes 
                         \OP\circ\gprov_\ProgD{}{\D}(\mskol{g^2_1}{\bar x}) \\
  \end{array}
\]
Since the program is positive, each goal node has exactly one following negated relation node.
Those negated relation nodes in turn have exactly one corresponding positive relation node.
As shown above for EDB facts and for positive programs with a head node $J(\bar x)\in \Q(\D)$,
positive relation nodes lead to the corresponding provenance semiring annotations:
\[
  \begin{array}{rcl}
     \OP\circ\gprov_\ProgD{}{\D}(\mskol{J}{\bar x})
                   & = & \OP\circ\gprov_\ProgD{}{\D}(\mskol{\neg R_1}{\bar x}) \KTimes 
                         \OP\circ\gprov_\ProgD{}{\D}(\mskol{\neg R_2}{\bar x}) \\
                   & = & \OP\circ\gprov_\ProgD{}{\D}(\mskol{R_1}{\bar x}) \KTimes
                         \OP\circ\gprov_\ProgD{}{\D}(\mskol{R_2}{\bar x}) \\
                   & = & \tj^\NX_\ProgD{}{\D}(R_1(\bar x)) \KTimes \tj^\NX_\ProgD{}{\D}(R_2(\bar x)) \\
  \end{array}
\]
\qed
\end{Proof}

\end{document}